\documentclass{emulateapj}

\usepackage{epsfig}

\newcommand{\uyvol}{UY~Vol}
\newcommand{\HST}{{\it HST}}

\newcommand{\RXTE}{{\it RXTE}}

\newcommand{\hst}{{\it HST}}
\newcommand{\xte}{{\it RXTE}}

\newcommand{\ebv}{E(B-V)}

\slugcomment{Accepted for publication in the Astrophysical Journal}

\shorttitle{Reprocessing of X-ray Bursts 
  in EXO~0748--676}
\shortauthors{Hynes et al.}

\begin{document}


\title{Multiwavelength Observations of EXO~0748--676 --
  I. Reprocessing of X-ray Bursts}

\author{R. I. Hynes\altaffilmark{1}, 
        Keith Horne\altaffilmark{2}, 
        K. O'Brien\altaffilmark{3}, 
        C. A. Haswell\altaffilmark{4},
        E. L. Robinson\altaffilmark{5}
        A. R. King\altaffilmark{6}
        P. A. Charles\altaffilmark{7,8},
        K. J. Pearson\altaffilmark{1}}
\altaffiltext{1}{Department of Physics and Astronomy, Louisiana State
University, Baton Rouge, Louisiana 70803, USA; rih@phys.lsu.edu}
\altaffiltext{2}{School of Physics and Astronomy, The University of St
        Andrews, St Andrews, KY16 9SS, UK}
\altaffiltext{3}{European Southern Observatory, Casilla 19001,
        Santiago 19, Chile}
\altaffiltext{4}{Department of Physics and Astronomy, The Open
        University, Walton Hall, Milton Keynes, MK7 6AA, UK}
\altaffiltext{5}{Astronomy Department and McDonald Observatory, The University of Texas at Austin, 
1 University Station C1400, Austin, Texas 78712, USA}
\altaffiltext{6}{Department of Physics and Astronomy, 
The University of Leicester, University Road, Leicester, LE1 7RH, UK}
\altaffiltext{7}{Department of Physics and Astronomy, 
The University of Southampton, Southampton, SO17 1BJ, UK}
\altaffiltext{8}{South African Astronomical Observatory
P.O. Box 9, Observatory, 7935, South Africa}

\begin{abstract}
  We present the first high time-resolution simultaneous X-ray,
  ultraviolet, and optical observations of X-ray bursts in \uyvol, the
  optical counterpart of the low mass X-ray binary EXO~0748--676,
  obtained with \RXTE, \HST, and Gemini-S.  Strong reprocessed signals
  are present in the ultraviolet (a factor of 4) and optical (a factor
  of 2.5).  These signals are lagged with respect to the X-rays and
  appear significantly smeared as well.  The addition of
  far-ultraviolet coverage for one burst, together with the high
  quality of the dataset, allow much tighter constraints upon the
  temperature and geometry of the reprocessing region than previously
  possible.  A single-zone black body reprocessing model for this
  burst suggests a rise in temperatures during the burst from
  18\,000 to 35\,000\,K and an emitting area comparable to that expected
  for the disk and/or irradiated companion star.  The lags, a mean of
  4.0\,s and range of 2.5\,s, are consistent with those expected
  within the binary.  The single-zone black body model cannot
  reproduce the ratio of optical to ultraviolet flux during the burst,
  however.  The discrepancy, corresponding to underpredicting the
  optical by more than a factor of two, seems too large to explain
  with deviations from a local black body spectrum and more likely indicates
  that a range of reprocessing temperatures are required, as would be
  expected, with cooler regions not contributing to the UV.
  Comparable results are derived from other bursts, and in particular
  the lag and smearing both appear shorter when the companion star is
  on the near side of the disk as predicted.  The burst observed by
  \HST\ also yielded a spectrum of the reprocessed light.  It is
  dominated by continuum, with a spectral shape consistent with the
  temperatures derived from lightcurve modeling.  Some line
  enhancements are also seen, most prominently in C\,{\sc iii}
  1175\,\AA.  Taken as a whole, our observations confirm the standard
  paradigm of prompt reprocessing distributed across the disk and
  companion star, with the response dominated by a thermalized
  continuum rather than by emission lines.
\end{abstract}


\section{Introduction}

The low-mass X-ray binary (LMXB) EXO~0748--676 was discovered in 1985
as a transient X-ray source \citep{Parmar:1985a} and rapidly
associated with an optical counterpart, \uyvol\ \citep{Wade:1985a}.
Unlike most X-ray transients, however, it did not decay back to a
quiescent state, but remained active and is now considered part of the
persistent LMXB population.  This makes it an intriguing object for
study as it appears to exist near the edge of stability, flipping
between phases of quasi-stable activity and quiescence.  The likely
key to its behavior is that the system is currently held in a
meta-stable high state by i) stabilization of the accretion disk
against thermal instability by X-ray irradiation and ii) enhancement
of the mass transfer rate from the companion.  \uyvol\ is thus an
important object for understanding the impact of X-ray irradiation
upon the disk and companion star in LMXBs.

\uyvol\ also has an inclination that is just right to give us a range
of diagnostic tools to probe the accretion flow.  The inclination is
high enough that X-ray eclipses are seen recurring on a 3.82\,hr
period, and X-ray dips also occur \citep{Parmar:1986a}.  The eclipses
are sharp, indicating that the neutron star itself is being eclipsed,
and hence that it is visible outside of eclipse.  \uyvol\ is therefore
not an accretion disk corona (ADC) source; an ADC is likely present,
but is not the dominant source of observed X-rays.

To better understand the accretion structure and effect of irradiation
in this LMXB, we have performed a multiwavelength study using \HST\
for UV rapid spectroscopy, \RXTE\ for X-ray data, Gemini-S for rapid
optical photometry, and the Cerro Tololo 4\,m Blanco telescope for
optical spectroscopy.  We discuss here the analysis of several X-ray
bursts seen during the simultaneous coverage.  Subsequent papers will
address the UV and optical emission line spectra
(\citealt{Pearson:2006a}, hereafter Paper II), and the spectral energy
distribution, and multiwavelength orbital light curves (Paper III).

Type I X-ray bursts, seen in low-mass X-ray binaries (LMXBs) with
neutron star primaries, are due to explosive thermonuclear burning of
accreted material on the surface of the neutron star.  They involve a
large increase in the X-ray flux, by a factor of ten or more, on
timescales of a few seconds.  As well as providing insights into the
conditions on the surface of the neutron star, the sudden flash lights
up the whole binary system, and significant reprocessed bursts, with
an amplitude of a factor of a few, are seen in the optical (see
\citealt{Hynes:2004a} and references therein).  Since the reprocessed
bursts are a large amplitude signal rather than a small perturbation,
the non-linearity of the optical response (due to observing
band-limited rather than bolometric fluxes) is clear.  Thus
lightcurves in different bandpasses exhibit significant differences,
breaking the degeneracy between the temperature and emitting area of
the reprocessing region that exists when only small perturbations are
considered \citep{Lawrence:1983a}.  This makes the bursts a very
powerful tool for applying echo mapping techniques in an X-ray binary
(e.g.\ \citealt{OBrien:2002a}; \citealt{Hynes:2004a}).

In spite of this potential, the unpredictable nature of the bursts
means that relatively few simultaneous observations exist of any X-ray
bursting source, many twenty years old, and no UV observations of
bursts have been obtained.  We present here the first simultaneous
X-ray, UV, and optical observations of a burst in \uyvol, including
fast far-UV spectroscopy.  We also present several more bursts
observed with only X-ray and optical coverage.

\section{Observations}

\begin{table}
\caption{Log of UV, optical, and X-ray high time resolution observations.}
\label{OpticalTable}
\begin{tabular}{lllcr}
\hline
\noalign{\smallskip}
Facility & Instrumentation & Start date & UT range & Total \\
         &                 &            &          & time (s) \\
\noalign{\smallskip}
\hline
\noalign{\smallskip}
\HST & STIS, G140L & 2003 Feb 18--19 & 20:06--00:06 & 13470 \\
\HST & STIS, G230L & 2003 Feb 19 & 00:17--00:31 &  800 \\
\HST & STIS, G140L & 2003 Feb 19 & 00:53--04:52 & 13470 \\
\HST & STIS, G230L & 2003 Feb 19 & 05:04--05:17 &  800 \\
\noalign{\smallskip}
\hline
\noalign{\smallskip}
Gemini-S & AcqCam, $V$ & 2003 Feb 18 & 04:19--05:32 & 9600 \\
Gemini-S & AcqCam, $V$ & 2003 Feb 19 & 01:08--08:38 & 18960 \\
\noalign{\smallskip}
\hline
\noalign{\smallskip}
\RXTE & PCA & 2003 Feb 14     & 00:08--09:32 & 10300 \\
\RXTE & PCA & 2003 Feb 15     & 01:27--09:12 &  8238 \\
\RXTE & PCA & 2003 Feb 17     & 17:21--18:23 &  3520 \\
\RXTE & PCA & 2003 Feb 17--18 & 20:30--08:13 & 11751 \\
\RXTE & PCA & 2003 Feb 18     & 12:17--13:19 &  1792 \\
\RXTE & PCA & 2003 Feb 18--19 & 20:11--09:23 & 13706 \\
\RXTE & PCA & 2003 Feb 19     & 15:07--16:09 &  3616 \\
\noalign{\smallskip}
\hline
\end{tabular}
\end{table}

\subsection{HST}
\HST\ observations used the Space Telescope Imaging Spectrograph
(STIS; \citealt{Profitt:2002a}), with the far-UV MAMA detector and the
G140L grating, and are detailed in Table~\ref{OpticalTable}.  TIMETAG
mode yielded a stream of detected events, with 125\,$\mu$s precision,
which could be used to reconstruct spectra for any desired
time-interval, as well as high time-resolution lightcurves.  The
observations were timed such that the target was within the continuous
viewing zone (CVZ).  Consequently we were able to observe over about
9\,hrs with only small gaps for wavelength calibrations and mode
changes.  This covered two complete binary orbits.

To obtain a lightcurve, we extracted source counts from a 50\,pixel
wide window centered on the target and background from two similar
regions on either side.  Geo-coronal H\,{\sc i} Ly$\alpha$ and O\,{\sc
i} 1304\,\AA\ lines were masked out of both source and background
regions, as were the extreme ends of the detector.  It was necessary
to use this background subtraction procedure as the background was
found to be larger than the nominal global 7\,cnt\,s$^{-1}$ dark
current and time dependent.  This intermittent `glow' is not easily
modeled \citep{Profitt:2002a} so is best removed empirically.  We did
this by subtracting a polynomial fit to the background lightcurve.
Source counts out of burst were $\sim30-100$\,s$^{-1}$ and the
estimated background within the source window was $\la7$\,s$^{-1}$;
for comparison the nominal dark current should only be
$\sim0.3$\,s$^{-1}$ within this window.

We used 1\,s time-resolution to approximately match the Gemini-S
optical photometry.  As discussed by \citet{Hynes:2003a}, \HST/STIS
absolute timing accuracy is uncertain at a level of up to a few
seconds, so we did not attempt precise barycenter corrections and
allowed the zero point of the \HST\ timing to be a free parameter in
subsequent analysis.

Flux calibration of the lightcurves was derived from the average
spectra, and takes advantage of the reddening measurement
($\ebv=0.06\pm0.03$\footnote{Note that this is substantially below the
previously quoted value of 0.4 (\citealt{Schoembs:1990a};
\citealt{Liu:2001a}).  Since that value was based on shifting the
object to lie on the main-sequence in a color-color diagram, it is not
expected to be reliable for an LMXB}) possible with complementary
near-UV data (see Paper III).  For each G140L spectrum we compared the
observed counts as a function of wavelength with the dereddened fluxes
\citep{Fitzpatrick:1999a} to determine the effective sensitivity,
hence defining our far-UV bandpass.  We then used this sensitivity
function to derive weighted average wavelengths and fluxes.  These
averages were consistent to $\la1$\,\AA\ in wavelength, and
$\la1$\,\%\ in flux between the six G140L spectra.  The effective
wavelength derived (for our spectral shape) was 1388\,\AA.  The
uncertainty in the calibration of observed fluxes is 4\,\%\ (from the
documented flux calibration of low-resolution STIS/MAMA modes;
\citealt{Profitt:2002a}).  However, the dominant term is due to the
uncertainty in dereddening -- this introduces a $\pm25$\,percent
uncertainty in the absolute calibration of the far-UV lightcurves, and
$\pm16$\,percent in the relative calibration of the UV and optical
fluxes (since these errors are correlated).  If the extinction curve
differs from the assumed \citet{Fitzpatrick:1999a} form then the error
could be larger.

\subsection{Gemini-S}
On the nights of February 17--18 and 18--19, we used the Acquisition
Camera (AcqCam) on Gemini-South to obtain fast $V$ band optical
photometry (see Table~\ref{OpticalTable} for details).  Conditions
were photometric, with realized image quality on target mostly
$\sim0.9$\,arcsec, although this degraded to $\sim1.2$\,arcsec in the
latter part of the second night.  For practical data acquisition it
was necessary to break each night into series of (usually)
4000\,images at a time.  The series obtained, totaling 37\,000
images, are listed in Table~\ref{OpticalTable}.

One of the great strengths of this instrument for fast photometry is
the fast readout.  We used the camera windowed and binned ($2\times2$
giving $256\times256$ 0.24\,arcsec binned pixels).  Other data acquisition
modifications were made to further minimize the dead-time between
images.  This dead-time was not absolutely constant, but was usually
0.305--0.310\,s, with occasional (less than one per few hundred) glitches
to as high as 0.44\,s.  The precise time-stamps were taken from when
the image completed writing to disk, so the start time of the
subsequent image (relative to the first in a series) was known.  The
start time of the first image is taken from a GPS based clock.

Basic reductions including bias removal, subtraction of the
significant dark current, and flat-fielding were done using the {\sc
agreduce} script within {\sc iraf}\footnote{IRAF is distributed by the
National Optical Astronomy Observatories, which are operated by the
Association of Universities for Research in Astronomy, Inc., under
cooperative agreement with the National Science Foundation.}.  The
windowed field includes one star much brighter than \uyvol, which
provided a high fidelity reference star and two other comparisons (one
brighter, one fainter) which were used to verify the accuracy of the
results.  We experimented with both small aperture photometry using
{\sc iraf} and optimal photometry as implemented in the Starlink {\sc
photom} package (see \citealt{Naylor:1998a} for the algorithm this is
based on).  We found negligible difference between the two methods
provided the aperture (for unweighted photometry) was optimized.  We
opted to use the results from the optimal algorithm as this explicitly
adjusts the weighting per image and so should be more robust against
changes in seeing.  We used the comparison stars to verify that the
standard deviations of the resulting lightcurves are dominated by the
formal errors in the photometry.  The formal errors for \uyvol\ were
typically $\sim2.5$\,percent per 0.8\,sec exposure.

Absolute calibration was done with respect to two stars from SA\,104
(Landolt 1992) on the first night.  We derive magnitudes of $V=13.63$
for the reference star and 17.4--18.0 for \uyvol\ (excluding bursts).
The magnitude range derived for \uyvol\ is within the range observed
by earlier studies (e.g.\ 17.1--18.1; \citealt{vanParadijs:1988a}).
The orbital lightcurves will be considered in a subsequent paper.

Finally, we corrected the observed magnitudes for interstellar
extinction (assuming $E(B-V)=0.06$ based on the 2175\,\AA\ feature in
our \HST\ data) and converted to fluxes with the conversion constant
of \citet{Fukugita:1995a}.
\subsection{RXTE}
X-ray observations were obtained with the Rossi X-ray Timing Explorer
(\RXTE) in two blocks timed to coincide with optical spectroscopy
(Paper II), and the optical and UV observations described above.  A
log is presented in Table~\ref{OpticalTable}.

X-ray lightcurves of the bursts were recovered from a 64\,channel
event mode of the Proportional Counter Array (PCA).  See
\citet{Jahoda:2005a} for discussion of the current status of the PCA.
Since the source counts are high during a burst, and the timescale is
short, we used all PCUs which were switched on, including PCU0.  We
initially processed lightcurves from each PCU separately, however, to
check that no background flares were present in PCU0, before combining
them.  Background lightcurves (with 16\,s time-resolution) were
constructed from the L7/240 faint source combined models dated 2002
Feb 1.  This should allow the most reliable background subtraction
from the pre-and post-burst lightcurves.  Since the L7 rate is
modified for bright sources, however, this model lightcurve is
contaminated during the burst.  We therefore interpolated between the
pre- and post-burst modeled background rates to define the burst
background and subtracted this.  We note that in practice the
background subtraction has little impact on the results described here
as we perform an additional empirical subtraction of the persistent
(pre- and post-burst) flux.  Hence more refined background models
would not significantly affect our results.

Lightcurves were initially extracted for 2--5\,keV, 5--12\,keV and
12--60\,keV bandpasses.  The burst profiles varied significantly with
energy, so it is important to be careful in choosing an appropriate
lightcurve for a deconvolution.  We therefore also constructed a
2--20\,keV integrated flux lightcurve.  To do this, we estimated
approximate per-channel conversions from count rates to fluxes using
{\sc xspec} and appropriate response matrices.  We used a black body
fit to the average burst spectrum to define these conversions to
ensure that the relative weighting within each channel was
approximately correct.  We then applied these conversions to each
channel and summed the fluxes.  This will be somewhat noisier than a
straight sum of count rates, but represents our best estimate of the
evolution of the flux irradiating the disk and the companion star.
Based on the black body fits to the spectra, the 2--20\,keV bandpass
accounts for most of the burst flux ($\ga90$\,\%), so additional
bolometric corrections are ignored.
%
%
\section{System parameters}
Any attempt to quantitatively model observations of \uyvol\ requires
system parameters.  \uyvol\ does not yet benefit from dynamical
estimates, so we must use more indirect constraints.

\begin{figure}
\epsfig{angle=90,width=3.4in,file=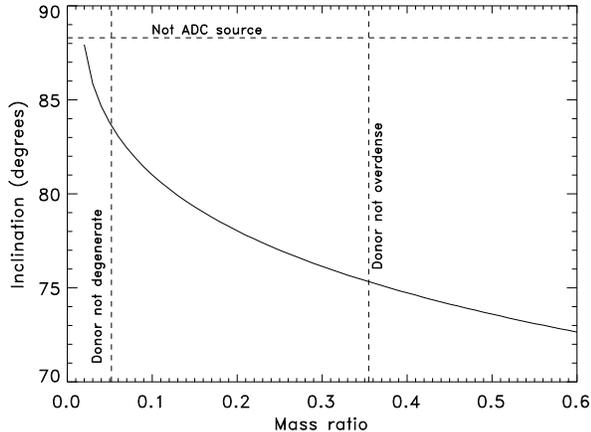}
\caption{Relation between inclination and mass ratio based on the
  widths of X-ray eclipses.}
\label{QvsIncFig}
\end{figure}

Fortunately it is eclipsing, and precisely defined X-ray eclipses
provide exquisitely detailed measurements of the orbital period and
duration of the neutron star eclipse \citep{Wolff:2002a}.  Through
these eclipses, we know that the inclination must be high, and the
relationship between mass ratio and inclination is well defined.
Assuming recent eclipse durations of $497.5\pm6.0$\,s, we obtain the
relationship shown in Fig.~\ref{QvsIncFig}.  We can attempt to further
constrain the available parameter space in other ways.  The fact that
we see the sharp neutron star eclipse at all indicates that \uyvol\
is not an accretion disk corona (ADC) source.  This means the
inclination cannot be too high as our line-of-sight must pass over the
disk rim.  The disk rim height is not known directly, but must be
greater than that expected from hydrostatic support alone.  This is a
very weak constraint, however; for example, for a disk half-thickness
of 0.03, we only require $i < 88.3^{\circ}$.  This constraint is shown
in Fig.~\ref{QvsIncFig}.

Other constraints are more model-dependent.  We can make plausible
estimates of the range of mass ratios likely, although values outside
of this range are still possible.  The companion star to \uyvol\ is
probably above the hydrogen burning limit and non-degenerate.  This
implies $M_2 \ga 0.07$\,M$_{\odot}$ \citep{Chabrier:2000a}.  Assuming
a 1.35\,M$_{\odot}$ neutron star \citep{Thorsett:1999a} then yields a
minimum mass ratio of 0.05.  This limit is soft, as a more massive
neutron star is certainly possible if it has accreted significant mass
from the companion (as seems likely in 2S~0921--630;
\citealt{Shahbaz:2004a}; \citealt{Jonker:2005a}).  At the high
mass-ratio end, we can assume that the companion star is underdense
compared to a main-sequence star that would fill the Roche lobe.
Again assuming a 1.35\,M$_{\odot}$ neutron star this yields the upper
limit on the mass ratio shown.  This could be increased a little for a
somewhat lower mass neutron star, and further if the companion is
actually overdense.  This is possible if the companion had undergone
nuclear evolution, before losing most of its envelope in mass transfer
(\citealt{Schenker:2002a}; \citealt{Haswell:2002a}).  In this case the
neutron star might also be expected to be more massive.  Consequently,
there is probably not much scope for a mass ratio higher than 0.4 in
this system.

It should be noted that the constraints considered so far are
effectively the same as obtained by \citet{Parmar:1986a}, namely
$75^{\circ} < i < 82^{\circ}$.  As discussed above, the limits are
rather soft, and a slightly larger range is possible with a
companion star and/or neutron star with extreme properties.

To obtain other parameters of interest, we synthesize a population of
possible binaries.  We assume the known orbital period, and the
relationship between mass ratio and inclination given in
Fig.~\ref{QvsIncFig}.  We consider neutron star masses with an
asymmetric Gaussian distribution $M_1 =
1.35^{+0.2}_{-0.04}$\,M$_{\odot}$, i.e., following
\citep{Thorsett:1999a}, but allowing for a higher mass tail due to
mass transfer.  We assume a uniform distribution of companion star
masses between 0.07\,M$_{\odot}$, and the mass at which main-sequence
density is reached.

We derive a relatively narrow range of binary separations,
$(1.03\pm0.05)\times 10^{11}$\,cm (at 90\,\%\ confidence), since the
binary period is known and there is not a large uncertainty in the
total system mass.  Disk parameters are more uncertain; the tidal
truncation radius is $(0.50\pm0.05)\times 10^{11}$\,cm and the
projected area of a flat disk would be $(1.5\pm0.2)\times
10^{21}$\,cm$^2$.  The projected area of the companion star, with a
spherical approximation, would be $(2.0\pm0.9)\times 10^{21}$\,cm$^2$.
While in general, the companion is expected to subtend a larger
projected area than the disk, only a phase-dependent fraction of this
will be X-ray heated, so this should be considered an upper limit on
the area of the luminous regions of the companion, which could be
significantly less than that of the disk.

We can estimate light travel times in the same way.  The maximum lag
from the pole of the companion star would be $(6.8\pm0.3)$\,s at phase 0.5,
although most of the heated inner face would of course be a little shorter
than this.  Lags from the disk would be expected to span a range from
zero to $(3.3\pm0.3)$\,s assuming it extends to the tidal truncation radius.

\begin{figure}
\epsfig{width=3.0in,file=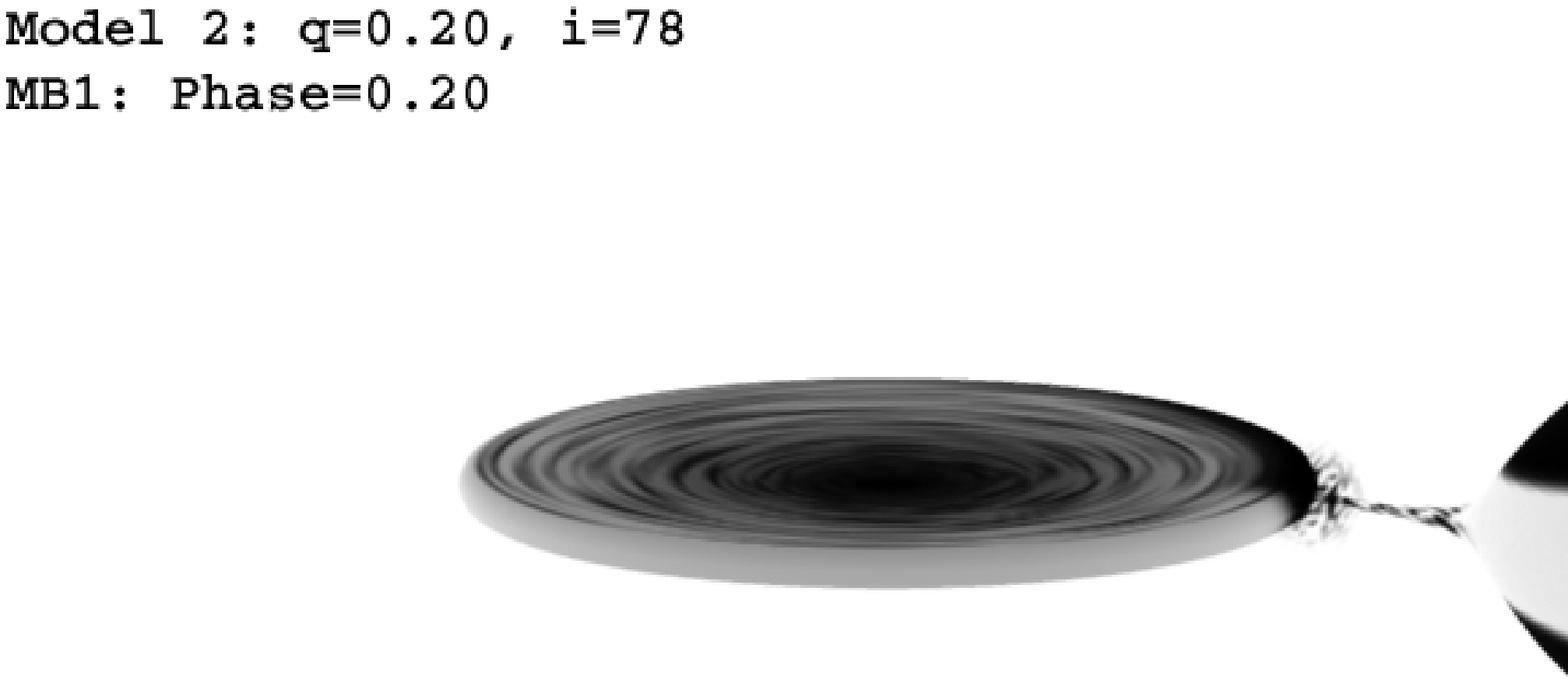}
\epsfig{width=3.0in,file=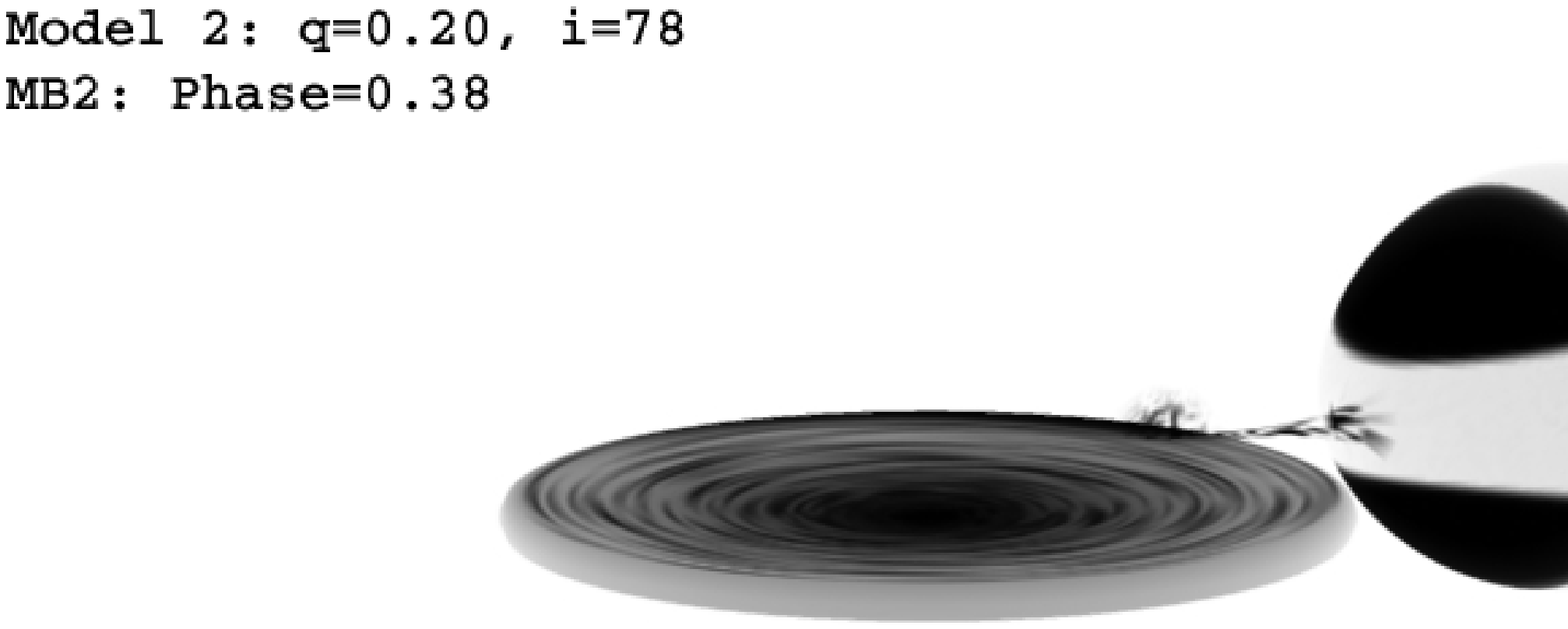}
\epsfig{width=3.0in,file=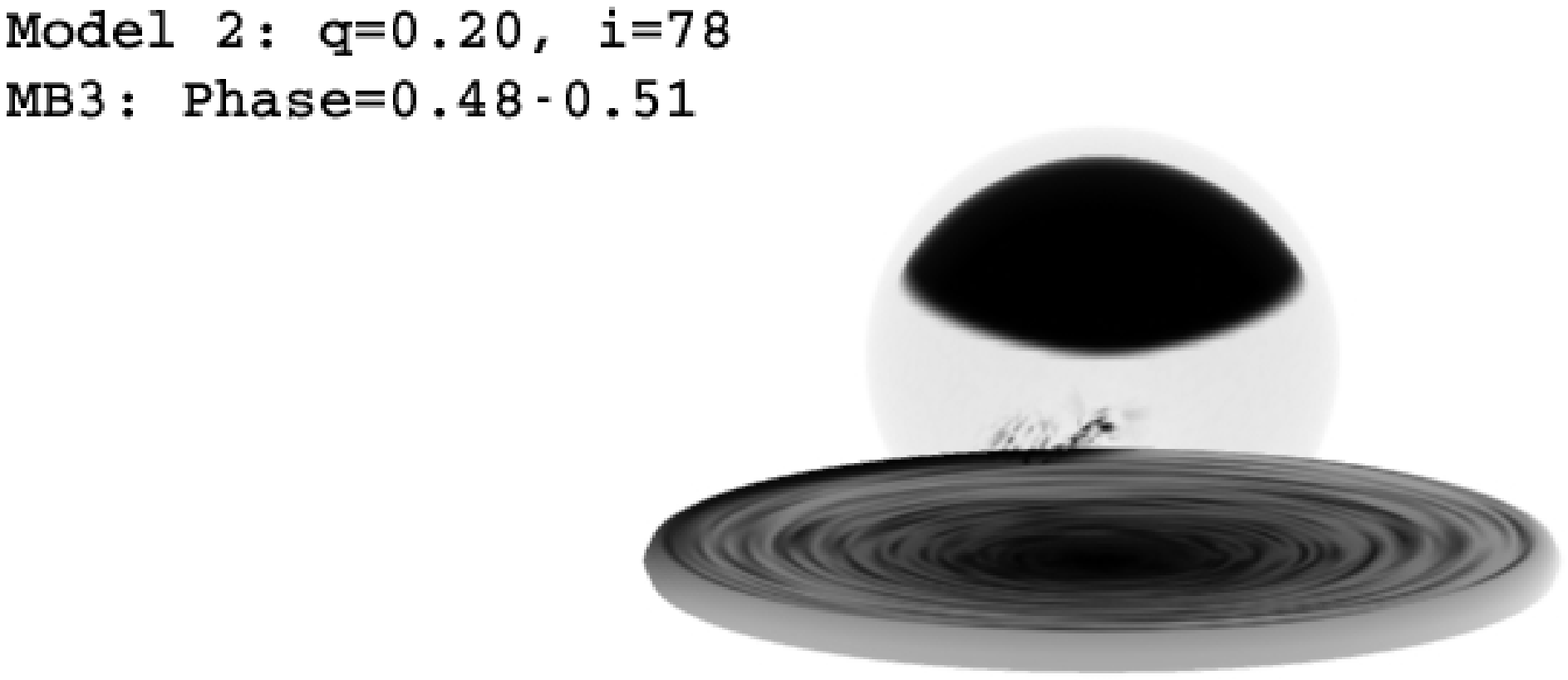}
\caption{System geometry at the phases where bursts were observed for
  model 2.}
\label{BinSimFig}
\end{figure}

We define several test cases consistent with Fig.~\ref{QvsIncFig}:
model 1 has $q=0.08$, $i=82^{\circ}$.  Model 2 has $q=0.2$,
$i=78^{\circ}$, and model 3 has $q=0.34$, $i=75.5^{\circ}$.
Fig.~\ref{BinSimFig} shows a schematic view corresponding to model 2
at phases when bursts were observed.

An important factor in considering irradiation of the companion star
is the opening angle of the disk, $\beta$.  For our purposes, this
defines the height of X-ray absorbing material, which may be above the
optical photosphere, and may not be in hydrostatic equilibrium (for
example material thrown up from the stream impact point or local
flares).  Values derived from other objects have generally been rather
high.  \citet{deJong:1996a} derived $\beta=12^{\circ}$ and cite other
authors who obtained a range of 6--14$^{\circ}$.  In our case, the
highest values are ruled out in some models, as \uyvol\ is not an ADC
source -- we see neutron star eclipses and prominent bursts, hence we
do observe the neutron star directly.  Thus the opening angle must be
less than $90-i$, i.e.\ less than $12^{\circ}$ for model 1 and less
than $8^{\circ}$ for model 2.  This is not a constraint for model 3.
Note that the presence of eclipses also indicates that at least some
of the companion star must be exposed to direct radiation from the
neutron star; it cannot be fully shielded by the disk.

A final important parameter is the distance to \uyvol.  The best
indicator of this is in neutron star LMXBs is the peak flux observed
during radius expansion X-ray bursts as this is believed to be an
approximate standard candle \citep{Kuulkers:2003a}.  Three radius
expansion bursts were reported by \citet{Gottwald:1986a} and
\citet{Jonker:2004a} used these, together with the calibration of
\citep{Kuulkers:2003a}, to derive a distance range to \uyvol\ of
6.8--9.1\,kpc, with the low end of the range corresponding to burning
of material of normal composition and the upper end to hydrogen poor
material.  More recently, \citet{Wolff:2005a} used a brighter radius expansion
burst seen by \RXTE\ to obtain distances of $5.9\pm0.9$\,kpc or
$7.7\pm0.9$\,kpc for hydrogen rich and poor bursts respectively.  The
latter authors argue that the variation in brightness likely reflects
variable obscuration and hence that their brighter burst gives a more
reliable distance measurement.  We
will thus use the midpoint of the latter estimates for this work,
6.8\,kpc.  Allowing for the composition ambiguity, all luminosities
and estimated areas quoted in this work will then be uncertain by
$\pm50$\,\% due to the uncertain distance.

%
\section{Bursts detected}

\begin{figure}
\epsfig{width=3.6in,file=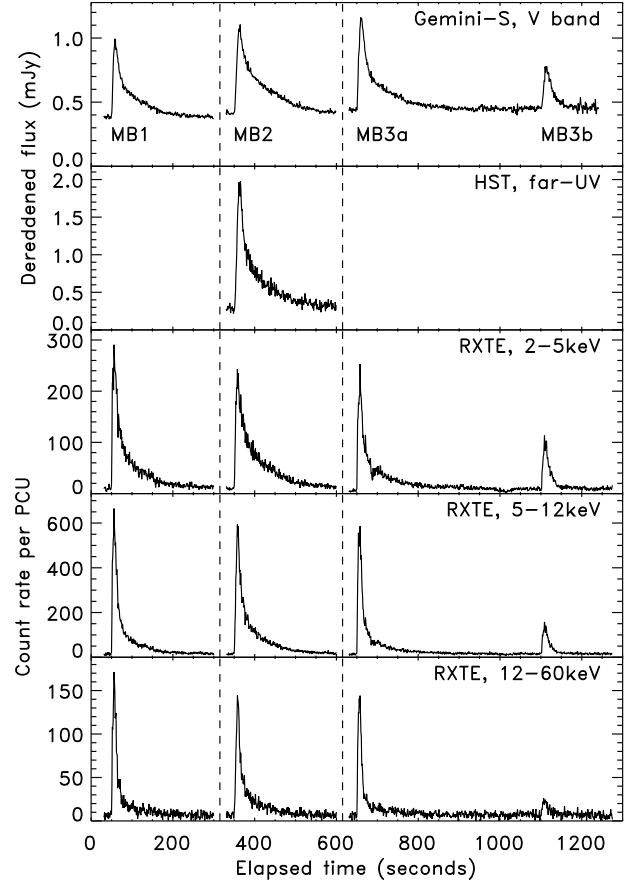}
\caption{X-ray bursts with simultaneous multiwavelength coverage.
  These have been arbitrarily shifted in time to fit on common axes.
  Dashed lines indicate non-contiguous coverage.}
\label{MultiBurstFig}
\end{figure}

Our \xte\ coverage detected a total of 6 bursts, including two weak
ones, listed in Table~\ref{BurstTable}.  One of these, MB3b, was
observed in a pair where the second is weaker than the first, and the
other, XB1, could also have been preceded by an unobserved normal
burst as it occurs at the beginning of a time-series.  Four of these
bursts, including the pair, were observed by Gemini-S and one of these
also by \hst, giving X-ray, UV, and optical coverage of the same
burst.  The data quality in the UV, and especially the optical, is
superb, making these the best observed optical bursts in any source.
The bursts with multiwavelength coverage are shown in Fig.~\ref{MultiBurstFig}.

\begin{table}
\caption{Bursts detected during our observations}
\label{BurstTable}
\begin{tabular}{llll}
\hline
\noalign{\smallskip}
Burst ID & Wavelengths & UT date and time & Phase \\
\noalign{\smallskip}
\hline
\noalign{\smallskip}
XB1  & X-ray           & 2003 Feb 15, 05:27:06 & 0.22 \\
XB2  & X-ray           & 2003 Feb 18, 01:15:16 & 0.95 \\
MB1  & X-ray, opt.     & 2003 Feb 18, 06:01:38 & 0.20 \\
MB2  & X-ray, opt., UV & 2003 Feb 19, 01:49:12 & 0.38 \\
MB3a & X-ray, opt.     & 2003 Feb 19, 06:01:36 & 0.48 \\
MB3b & X-ray, opt.     & 2003 Feb 19, 06:09:08 & 0.51 \\
\noalign{\smallskip}
\hline
\end{tabular}
\end{table}

One of the simultaneously observed bursts, MB3a, shows dips in the
otherwise smooth decay.  This burst occurred at orbital phase 0.478, so
does not lie within the classical dipping phase-range but our X-ray
data do appear to show additional dipping near phase 0.5 (Paper III),
so it might be due to transient absorption by intervening material.
In support of this, the dip does appear most pronounced in the
2--5\,keV energy band, and is absent in the optical data.  Both these
characteristics suggest that the dips are due to absorption of the
direct X-rays along our line-of-sight, rather than variations in the
intrinsic burst luminosity.

%
\section{Burst Analysis}
\subsection{Cross-correlation Functions}

\begin{figure*}
\epsfig{angle=90,width=6.8in,file=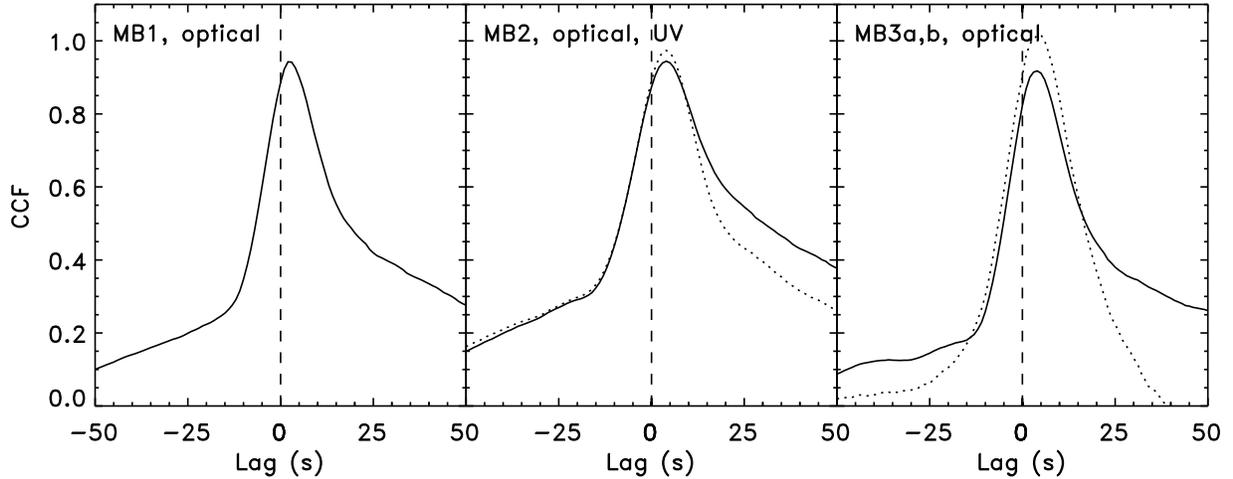}
\caption{Cross-correlation functions for the bursts.  The optical or
  UV broad-band data are cross-correlated against the integrated X-ray
  flux lightcurves.  For MB2, the solid line shows the result with
  optical data, the dotted one that with UV.  For MB3, the solid lines
  shows MB3a, the dotted one MB3b.}
\label{CCFFig}
\end{figure*}

Cross-correlation functions (CCFs) provide a relatively simple
technique for examining lags between lightcurves, and one which is
widely used allowing comparison with other datasets.  We therefore
begin with this approach, before developing other techniques
customized to the reprocessed burst problem.  The datasets we have are
almost exactly evenly sampled, but the Gemini-S photometry has a
somewhat longer sampling interval than the \xte\ or \hst\ data.
Interpolation cross-correlation functions (\citealt{Gaskell:1987a};
\citealt{White:1994a}) are thus ideally suited to these data.  We show
all of the CCFs in Fig.~\ref{CCFFig}.  The single, strongly correlated
signal makes identification of the peak very easy, and its location is
well defined.  A lag of a few seconds is clearly seen, with individual
values tabulated in Table~\ref{BurstFitTable}.  All lags are of order
a few seconds, as expected for light travel times within the binary.
All full strength bursts show rather similar structure, and asymmetry
between the rising and decaying portions.  For MB3a and MB3b, the lags
derived are consistent, but the shapes of the CCF are not.  The CCF
lags do suggest a positive correlation between orbital phase and mean
lag, but the evidence is not compelling (and would vanish if MB1 were
removed from the sample).  Nonetheless, this is as expected if there
is a significant contribution to the reprocessed signal from the
surface of the companion star, as the lag from this will be minimum at
phase 0.0 and maximum at phase 0.5.  This behavior is also consistent
with the finding of \citet{Schoembs:1990a} that the {\em rise time} of
optical bursts (without simultaneous X-ray coverage) was correlated
with the orbital phase in the same sense.

\begin{table}
\caption{Lags measured from CCF.}
\label{BurstFitTable}
\begin{tabular}{llll}
\hline
\noalign{\smallskip}
Burst ID & Wavelengths & CCF Lag & Phase \\
\noalign{\smallskip}
\hline
\noalign{\smallskip}
MB1  & X-ray, opt. & 2.92 & 0.20 \\
MB2  & X-ray, opt. & 4.14 & 0.38 \\
MB2  & X-ray, UV   & 4.01 & 0.38 \\
MB3a & X-ray, opt. & 4.09 & 0.48 \\
MB3b & X-ray, opt. & 4.30 & 0.51 \\
\noalign{\smallskip}
\hline
\end{tabular}
\end{table}

Obviously before extensive interpretation, more sophisticated analysis
is warranted to verify these results.  The quality of the CCFs are
superb and the major concern relates to the meaningfulness of the lag
derived.  Fundamental to the CCF approach is the assumption that the
optical/UV is simply a lagged version of the X-rays, but this is
clearly not the case in the burst lightcurves, and is not expected
theoretically.  Several factors may be important.  One is light travel
times.  These will not only produce a mean lag, but will tend to smear
out the signal, as different lags are produced by different
reprocessing sites \citep{OBrien:2002a}.  A second factor having a
similar effect is that there may be a finite diffusion time associated
with reprocessing; if X-rays deposit energy at a significant optical
depth it cannot be re-radiated instantly, but there will be further
lagging and smearing as the energy diffuses outward.  Finally, the
tails of the bursts are clearly different, indicating non-linear
relationships between the lightcurves.  This occurs because the bursts
have a large amplitude and substantially change the temperature of
both the neutron star and the reprocessing site.  As a site cools, the
peak of the emitted spectrum moves to lower energies.  In the case of
the reprocessed light, it moves out of the far-UV, and into the
optical, having the effect of accelerating the far-UV decay rate
(relative to the bolometric decay) and suppressing the optical decay.
None of these factors are accounted for in cross-correlations, so a
more sophisticated model is needed.  For example, lags can be over or
under-estimated if the timing characteristics of the two lightcurves
differ \citep{Koen:2003a}, reflecting the more general problem that in
this case there is not a uniquely defined lag.  There is a range, from
which a CCF favors an average value.

\subsection{Maximum Entropy Deconvolutions}

We next attempt to deconvolve the lightcurves to obtain the transfer
function between X-ray and optical wavelengths independent of the
width of the burst.  This will be more readily interpreted than the
CCF which is much broader than the transfer function.  We initially
use the maximum entropy method (MEM; \citealt{Horne:1991a};
\citealt{Horne:1994a}), which does not require an assumed model for
the transfer function, beyond an assumed default used in calculating
entropy.  In principle, the MEM method can then resolve distinct
sub-structures, if they are present, for example the disk and
companion star.

\begin{figure*}
\epsfig{angle=90,width=6.8in,file=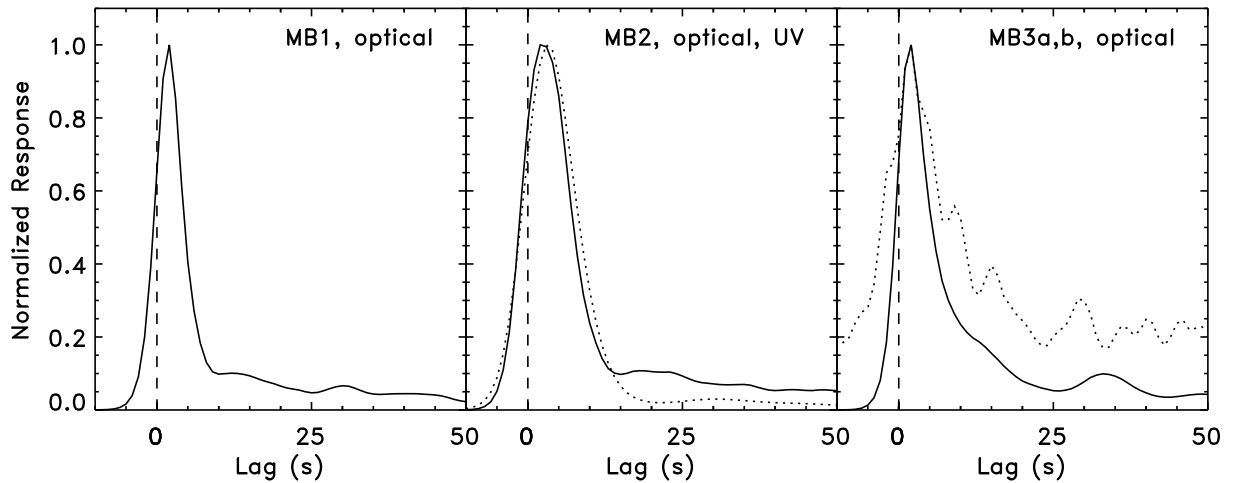}
\caption{MEM reconstructed transfer functions for the simultaneous
  bursts.  For MB2, the solid line shows the result with optical data,
  the dotted one that with UV.  For MB3, the solid lines shows MB3a,
  the dotted one MB3b.}
\label{MEMFig}
\end{figure*}

For each burst we use our X-ray flux lightcurves as the driver signal
and attempt to fit the optical or UV echo over the range $-50$ to
$+300$\,s relative to the burst rise.  The transfer function was
calculated for lags of $-50$\,s to $+200$\,s.  We use a narrow
Gaussian default to minimize enforced smoothing of the derived
transfer functions.  The results are shown in Fig.~\ref{MEMFig}.
Negligible response was seen before $-10$\,s, and slow declines after
$+50$\,s; these regions are not shown.

As expected, the derived transfer functions are much narrower than the
CCFs.  All bursts suggest a single-peaked response peaking at a lag of
a few seconds, very close to the locations of the maxima of the
corresponding CCFs.  Besides the main peak, all the transfer functions
have a tail extending to larger lags.  This may be a real effect, as
models of reprocessing of bursts by stellar atmospheres do predict a
small amount of the energy emerges with large delays
(\citealt{Cominsky:1987a}; see Section~\ref{DiffusionSection} for more
discussion).  We cannot say with confidence that this effect is real,
however, as there is another explanation for it.  As noted earlier,
the bandpass-limited response is non-linear, and in particular the
optical burst decay is much slower than that in X-rays because the
reprocessed spectrum shifts to longer wavelengths (into the optical
band) as it cools.  A method which assumes a linear response will
require an extended tail to reproduce this.  In support of this
interpretation, the effect is much weaker in the transfer function
derived from UV data as expected.

This effect complicates the interpretation of the shape of the
transfer functions using this method.  In addition, while it is
tempting to ascribe the narrower transfer function from MB1 to a
phase-dependent effect, the response to MB3a is also narrower, and in
general one has to be cautious in interpreting widths derived from MEM
echo maps as a broader, smoother peak inherently has higher entropy.
We therefore seek a less biased approach.

\subsection{Gaussian Transfer Function Fitting}
\subsubsection{Method}
\citet{Hynes:1998b} demonstrated a method for constraining the
transfer function by parameterizing it as a Gaussian function.  Our
MEM reconstructions indicate that the data do not require multiple
resolved components in the transfer function.  This does not mean that
the true transfer function does not include multiple components; only
that they are not well constrained by the data.  This is probably a
consequence of the relatively long burst timescale, which limits our
sensitivity to fine temporal structure in the response.  Consequently
the single Gaussian transfer function is an adequate approximation to
this problem and it lends itself well to adapting to the rather
different assumptions needed for burst mapping.

These arise because of the large amplitude of the variations.  We have
already discussed one problem this causes; the X-ray lightcurves are
strongly energy dependent, so determining the driving lightcurve is
non-trivial.  This arises because the X-ray flux arises from heating
and cooling of the neutron star surface; as the surface cools, the
spectrum softens and moves out of the higher energy bands.  We have
addressed this by calculating a 2--20\,keV integrated flux lightcurve,
using the spectral information to ensure the correct internal relative
weighting.  Similar difficulties exist with the optical and UV data.
The optical burst is more prolonged than the UV one, for the same
reason as the bursts are more prolonged at low energies; the
reprocessed spectrum is moving out of the UV band into the optical as
the reprocessing regions cool.  The net effect of this is that the
optical and UV fluxes do not respond linearly to the X-ray flux, and
the amplitude of the variations is too large for them to be treated as
a small, linearized perturbation.  This is actually a blessing in
disguise however, as we can then use the relative changes between the
optical and UV fluxes to constrain the temperature evolution, and
hence also the emitting area of the reprocessing region.  Without this
information, there would be a degeneracy between temperature and area,
and indeed this is a problem when we try to fit optical or UV data
independently.  For example, \citet{Lawrence:1983a} attempted a
similar analysis of 4U\,1636--536 using $UBV$ photometry and were able
to estimate that the bursts involved a rise in temperature of the
reprocessor from $\sim25,000$\,K to $\sim50,000$\,K.

We formulate the problem as follows.  We assume that an input X-ray
lightcurve (the 2--20\,keV integrated flux) is a suitable proxy for
the bolometric illuminating luminosity, $L_{\rm X}$.  Reprocessing is
assumed to occur within a fixed area $A$ (assumed to be of negligible
geometric depth) with uniform temperature $T(t)$.  The spatial
position and extent of this region introduces a mean lag, $\tau$, and
a blur $\sigma_{\tau}$ in the reprocessed bolometric luminosity
relative to the illuminating luminosity.  The temperature is then
allowed to vary (with fixed reprocessing area) such that $T^4(t)
\propto L_{\rm rep}$, where the reprocessed luminosity is assumed to
be related to the X-ray luminosity via the transfer function, $\Psi$:
$L_{\rm rep} \propto \Psi * L_{\rm X}$.  Given the temperature
evolution and area we can then predict the bandpass-limited UV and
optical flux evolution.  This requires a model for the reprocessed
spectrum, for which we assume a black body in the absence of a better
choice.  The problem can be parameterized in terms of $\tau$ and
$\sigma_{\tau}$, defining the transfer function (which is arbitrarily
normalized), $T_{\rm min}$ and $T_{\rm irr}$ which are the minimum
temperature and the peak irradiation temperature, defined such that
$T^4_{\rm max} = T^4_{\rm min} + T^4_{\rm irr}$, and $A$, the
projected cross-sectional area of the reprocessing region.  Ideally
this gives enough parameters to fit the UV and optical lightcurves
simultaneously.  In practice, some additional nuisance parameters are
needed.  Since the \HST\ absolute timing is uncertain, we fit the
optical and UV lags independently; the optical should be reliable, but
the UV one is not, comprising both a true lag and a clock uncertainty.
We also allow the optical and UV bursts to have different reprocessing
areas.  This is initially assumed to reflect uncertainties in the
calibration of the optical photometry and in dereddening of the
optical:UV flux ratio, rather than physical differences.  In the ideal
case, both lags and both areas would be equal.

\subsubsection{The multiwavelength burst (MB2)}
We begin with the burst which has both UV and optical coverage as this
should be best constrained.  It also allows us to test the reliability
of the optical fit by comparing it with the joint optical-UV fit.

\begin{figure}
\epsfig{width=3.6in,file=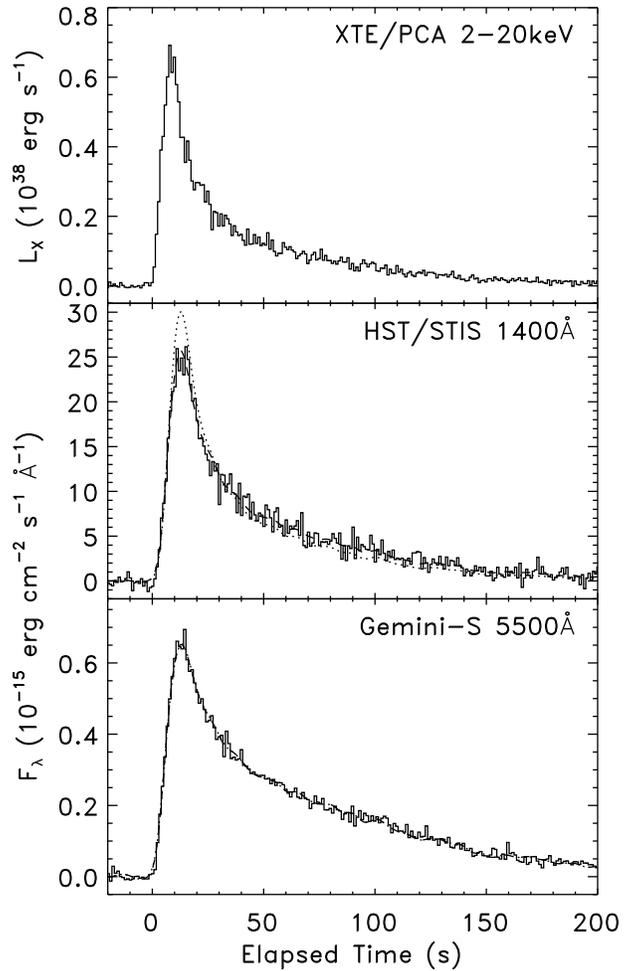}
\caption{Single black body fits to MB2, allowing the optical-to-UV areas
to float independently (dashed line), or fixing them to the same value
(dotted line).  The underlying non-burst X-ray luminosity that has
been subtracted off is $2.1\times10^{36}$\,erg\,s$^{-1}$.}
\label{MBFirstFig}
\end{figure}

We search for the best fitting models (in the $\chi^2$ sense) using a
downhill simplex method \citep{Press:1992a}.  To guard against
converging on a local minimum each fit used three passes, each
starting at the previous best fit, and we also varied the initial
starting simplex to approach the minimum from different directions in
parameter space.  We find that the method can reproduce both the
optical and UV lightcurves very well.  The best fitting joint model
(MB2-1) is shown in Fig.~\ref{MBFirstFig} and
Table~\ref{GaussFitTable}, with approximately the same parameters
found for a variety of starting simplexes.  There was a small jitter
dependent on the starting point, but this was much less than the
uncertainties that we estimate, so all fits agreed to within errors.
The best fitting reduced $\chi^2$ is 1.85, but this is pessimistic.
The reason is that we have used the X-ray lightcurve to construct the
model, and its errors have not been accounted for.  To compensate for
this in deriving estimates of errors on the fit parameters, we use the
common technique of rescaling the errors so that the best fit model
has $\chi^2/{\rm dof} = 1$.

\begin{table*}
\caption{Fits to bursts with Gaussian transfer functions.}
\label{GaussFitTable}
\begin{tabular}{lllllllll}
\hline
\noalign{\smallskip}
Burst model & $\tau_{\rm opt}$ (s) & $\tau_{\rm uv}$ (s) & $\sigma_{\tau}$ (s) &
$T_{\rm min}$ (K) & $T_{\rm irr}$ (K) & 
$A_{\rm opt}$\tablenotemark{a} (cm$^2$) & $A_{\rm uv}$\tablenotemark{a} (cm$^2$) & $\chi^2/{\rm dof}$ \\

MB1-1 & $2.57\pm0.02$ & --            & $0.43\pm0.01$ & [18500]                  & $34800\pm500$   & $(1.88\pm0.03)\times 10^{21}$    & --           & 1.19 \\
MB1-2 & $2.44\pm0.04$ & -- & $1.23^{+0.06}_{-0.11}$   & [18500]                  & $35400\pm400$   & $(1.72\pm0.02)\times 10^{21}$    & --           & 1.33 \\
\noalign{\smallskip}
MB2-1 & $4.05\pm0.07$ & $3.80\pm0.18$ & $2.48\pm0.08$ & $18500\pm700$            & $35300\pm1200$ & $(1.92^{+0.09}_{-0.07})\times10^{21}$ & $(0.87^{+0.12}_{-0.09})\times10^{21}$ & 1.85 \\
MB2-2 & $4.12\pm0.08$ & $3.81\pm0.16$ & $2.62\pm0.09$ & $13600\pm100$\tablenotemark{b} & $26700\pm200$\tablenotemark{b} & $(2.65\pm0.05)\times10^{21}$ & --        & 2.12 \\
MB2-3 & $4.05\pm0.07$ & $3.80\pm0.17$ & $2.48\pm0.08$ & $12000\pm700$            & $23500\pm1200$ & $(2.32\pm0.01)\times10^{21}$ & $(1.31^{+0.09}_{-0.05})\times10^{21}$ & 1.85 \\
MB2-4 & $4.08\pm0.07$ & $3.81\pm0.16$ & $2.54\pm0.09$ & $9100\pm100$\tablenotemark{b} & $18700\pm100$\tablenotemark{b} & $(3.03\pm0.02)\times10^{21}$ & $(3.03\pm0.02)\times10^{21}$ & 1.90 \\
MB2-5 & --            & $3.78\pm0.15$ & $2.43\pm0.21$ & [18500]                  & $35900\pm1500$ & --                                    & $(0.82^{+0.12}_{-0.09})\times10^{21}$ & 1.11 \\
MB2-6 & $4.05\pm0.08$ & --            & $2.48\pm0.10$ & [18500]                  & $35300\pm400$  & $(1.92\pm0.05)\times10^{21}$          & --
& 2.69 \\
\noalign{\smallskip}
MB3-1 & $4.02\pm0.09$ & --            & $2.69\pm0.11$ & [18500]                  & $32700\pm500$  & $(2.50^{+0.03}_{-0.04})\times10^{21}$ & -- & 4.57 \\                    
\noalign{\smallskip}
\hline
\tablenotetext{a}{The errors quoted on areas only include statistical uncertainties; distance, reddening, and calibration uncertainties will dominate.}
\tablenotetext{b}{These uncertainties are unrealistically small.  In this model, the temperature is set precisely by the ratio of optical to UV flux, but the uncertainty only includes statistical errors.}
\end{tabular}
\end{table*}

The model derived has reasonable parameters.  The heated region has a
quiescent temperature of 18,500\,K and is heated to a peak temperature
of 36,000\,K.  The mean lag is 4.0\,s, and there is considerable
smearing, with a FWHM of 5.8\,s.  The implied error in the \HST\
timing is only 0.2\,s, which is better than might be expected.  The
lag is of the order expected for a combination of disk and companion
star reprocessing, but the smearing involved is large.  The inferred
emitting projected area is $0.9\times10^{21}$\,cm$^{2}$ from the UV
data and $1.9\times10^{21}$\,cm$^{2}$ using the optical.  The areas
derived are fully consistent with expectations from reprocessing on
the accretion disk ($1.5\pm0.2\times10^{21}$\,cm$^2$) and/or companion
star ($2.0\pm0.9\times10^{21}$\,cm$^2$).  The large discrepancy in the
area derived from the optical and UV is more problematic, however,
indicating that the optical/UV flux ratio is very different from that
implied by the temperatures; the optical is a factor of $\sim2$
brighter than expected.  This is much larger than can be accommodated
from errors in the dereddening (16\,\%), UV flux calibration (4\,\%)
or optical calibration (probably less than 10\,\%).  The difference is
clearly significant; a similar fit with the areas forced to be equal
(MB2-2) results in a visibly poor fit to the UV lightcurve.  Since the
derived magnitudes are already at the low end of the range observed by
previous studies it is more likely that the optical flux has been
underestimated than that it has been overestimated by a magnitude as
required.  If the extinction curve is far from the Galactic average,
then a larger reddening correction to the UV is possible, and this
could rectify the problem.

Alternatively, there may be a deficiency in the model.  The most
likely interpretation is that the reprocessing region is not
isothermal; indeed this is not expected to be the case.  The disk is
expected to span wide range of temperatures from rim to center, and
the companion star will provide an additional reprocessing site.  A
multi-temperature spectrum will be broader than a single temperature
one, and if the peak is in the UV (as implied by the derived
temperatures), then this will flatten the optical tail and increase
the optical fluxes.  A simple test of whether this explanation can
help resolve the discrepancy is to use simple irradiated disk spectra
as discussed by \citep{Hynes:2002a}.  Instead of using black bodies of
temperature $T$, we use irradiated disk spectra with irradiation
temperature at the edge $T$, and viscous temperature set to zero.  In
practice the latter constraint only means that the temperature is
dominated by irradiative heating; increasing the viscous temperature
reduced the quality of the fit.  Models MB2-3 and MB2-4 show parameters
derived using this alternate spectral model for floating and fixed
areas respectively.  Clearly this model does represent an improvement,
as the discrepancy between optical and UV areas has been reduced when
they are allowed to float, and the quality of fit is substantially
better if they are fixed to be the same.  This does not appear to be
the whole story, however, as a discrepancy remains.  This is probably
a consequence of the cooler response from the companion star (which is
inferred to contribute; see Section~\ref{EchoTomSection}).  We tried
adding another black body component, but found that beyond this point
fits were poorly constrained and multiple solutions were possible,
all yielding good fits with no discrepancy in optical and UV areas.  We
feel that a better way to approach this problem is to use a
model of the binary following the approach of \citet{OBrien:2002a}, in
which the binary geometry reduces the independence of the parameters
that we currently have using arbitrary components.  Such a model
would, however, benefit from information gleaned from other aspects of
our dataset, for example the orbital lightcurves, so we defer this
treatment to Paper III.

As a precursor to examining MB1 and MB3, for which only optical data
are available, we also tested fitting optical and UV lightcurves
separately.  We initially tried fitting with all the parameters left
free, but found that the problem was then very poorly constrained; a
single bandpass did not allow us to uniquely determine both minimum
and maximum temperatures.  The formal minimum for an optical-only fit
occurs for very low $T_{\rm min}\sim3500$\,K.  While this may be of
order the stellar temperature, the regions of the star exposed to
bursts are also exposed to persistent radiation, and should not be
this cool.  Furthermore, this solution is strongly inconsistent with
the UV lightcurve, and is only a shallow minimum; $\chi^2$ is
virtually constant for all $0 < T_{\rm min}< 50,000$\,K.  We thus
chose to instead fix $T_{\rm min}$ to the value determined by the
joint fit, 18,500\,K.  The parameters derived in this case (MB2-5 and
MB2-6) were not significantly different from those based on the joint
fit.  The lags cannot be reliably compared, due to the uncertainty in
\HST\ timing, but there is no significant difference between the
widths derived.

One important statement can be made from these analyses.
Independently of whether we fit optical and UV data jointly or
separately, and regardless of the spectral model assumed, we derive
essentially the same lags and widths, i.e.\ the transfer function
derived is robust and not sensitive to these assumptions.  This means
we can usefully compare transfer functions from different bursts that
do not have UV coverage, and without knowing the correct spectral
model to use.  

\subsubsection{The double burst (MB3)}

\begin{figure*}
\epsfig{width=6.4in,file=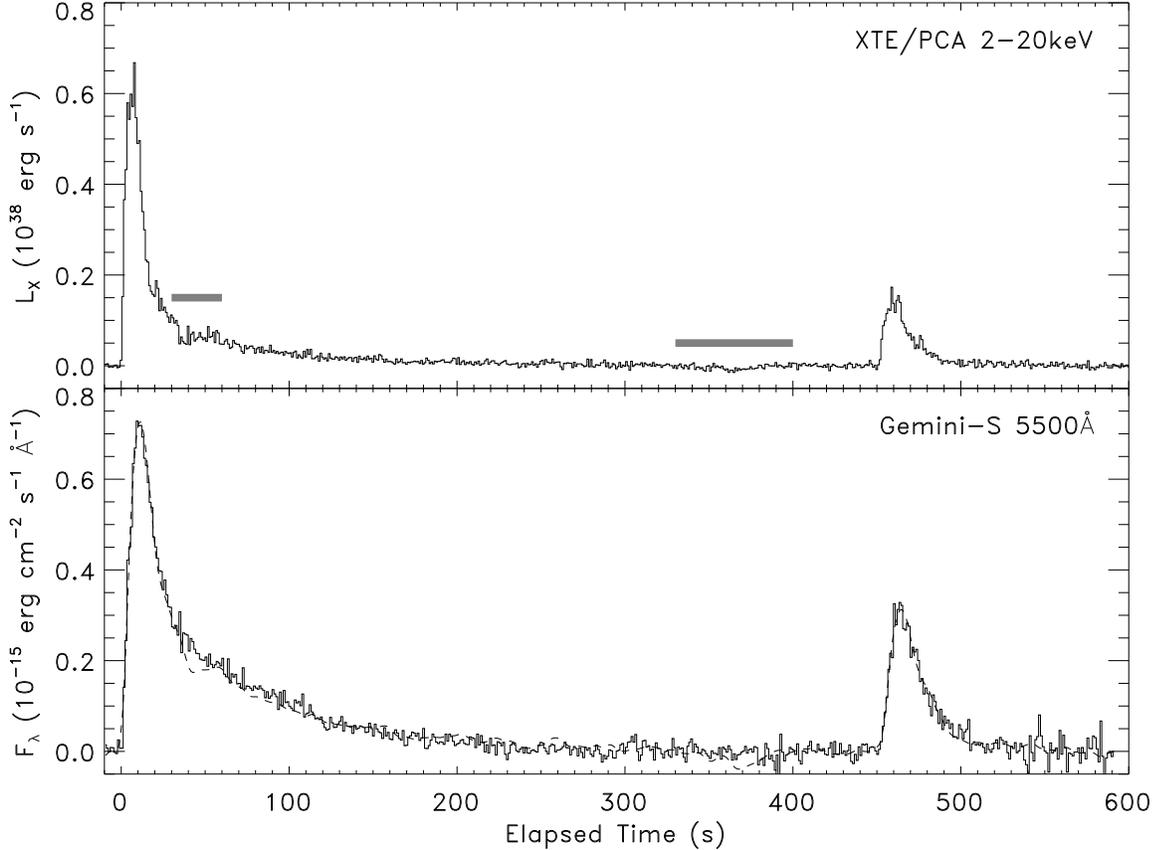}
\caption{Single black body joint-fit to both MB3a and MB3b.  The
underlying non-burst X-ray luminosity that has been subtracted off is
$1.9\times10^{36}$\,erg\,s$^{-1}$.  Gray bars indicate X-ray dipping
regions.  These were retained in calculating the model optical
lightcurve but were masked out of $\chi^2$ calculation.  The
discrepancy between the model and the optical data at these points is
further evidence that the dips are a consequence of our changing
visibility of the X-ray source, not a true change in the unabsorbed
luminosity.}
\label{MBThreeFig}
\end{figure*}

We now consider the double burst, MB3.  This is a more complex
case than MB1 but occurred only about four hours (one binary orbit) 
after MB2 on the same
night, so there are less likely to have been large changes in the
accretion flow geometry than for the first burst which occurred about
20\,hours (5 binary orbits) earlier.  Since MB3a and MB3b
are so close together, we perform a fit to both simultaneously, as
this will provide a visual check of the consistency between the two
bursts.  The major difficulty with these bursts is that they exhibit
dips, most prominently around 35\,s and 360\,s.  These dips are
clearly not present in the optical data, and appear more prominent at low
energies (Fig.~\ref{MultiBurstFig}).  This suggests that the dips are
due to absorption of the X-rays, and should be excluded from any
attempt to model the optical data.  We do this by masking out regions
of the optical lightcurve that would be affected.  The masked regions
are indicated in Fig.~\ref{MBThreeFig}.

As for MB2, $T_{\rm min}$ is not meaningfully constrained by the data
in the absence of UV coverage.  For $T_{\rm min}\ga6000$\,K, $\chi^2$
is virtually constant, and while cooler solutions are formally
favored, we choose to ignore them for the reasons described above.  A
fit with $T_{\rm min}=18,500$\,K is not visibly worse than one with
very low $T_{\rm min}$, so we choose to use this value for lack of
better information.  This will facilitate a more direct comparison
with MB2.

The fits to the lightcurves of both MB3a and MB3b are shown in
Fig.~\ref{MBThreeFig}, and parameters are given in
Table~\ref{GaussFitTable}.  Given that only $T_{\rm min}$ is externally
constrained, the agreement with MB2 is striking.  $\tau_{\rm opt}$
values agree within uncertainties and $\sigma_{\tau}$ virtually
agrees.  A slightly larger width is derived from MB3, but the
difference may not be significant.  It would be expected if the
response arises from a combination of disk and companion star,
however, as MB3 occurred around phase 0.5 when the lag from the
companion star is maximized.  The difference in temperatures and areas
is larger, but may be related to each other.  We have artificially
fixed $T_{\rm min}$, which will have the effect of also constraining
$T_{\rm max}$.  If $T_{\rm min}$ had actually increased between
bursts, then the corresponding $T_{\rm irr}$ would also have been
higher, and a lower area would have been required to fit the MB3
fluxes.  For example, a fit with $T_{\rm min}$ fixed to 21,000\,K
greatly reduces the discrepancy between $T_{\rm irr}$ and $A_{\rm opt}$
between MB2 and MB3.  Without UV data, however, we cannot say which of
these (or other) combinations are correct, so have chosen to assume
that $T_{\rm min}$ does not change.  

\subsubsection{The first burst (MB1)}
\begin{figure}
\epsfig{angle=90,width=3.5in,file=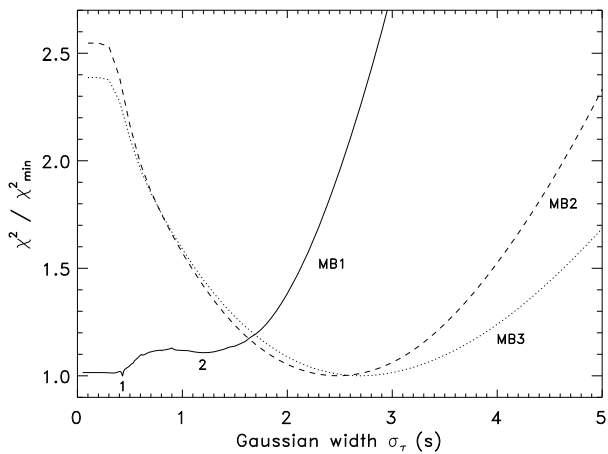}
\caption{Normalized $\chi^2$ as a function of Gaussian width for each
  burst.  For each point the width was fixed, but other parameters
  were optimized to find the best fit.  Annotations 1 and 2 for MB1
  refer to the two minima given as solutions in
  Table~\ref{GaussFitTable}.  Note that all curves flatten below
  $\sigma_{\tau}\sim0.3$\,s, corresponding to a Gaussian FWHM less
  than 0.7\,s, and less than the optical time-resolution.}
\label{WidthChiFig}
\end{figure}

\begin{figure}
\epsfig{width=3.5in,file=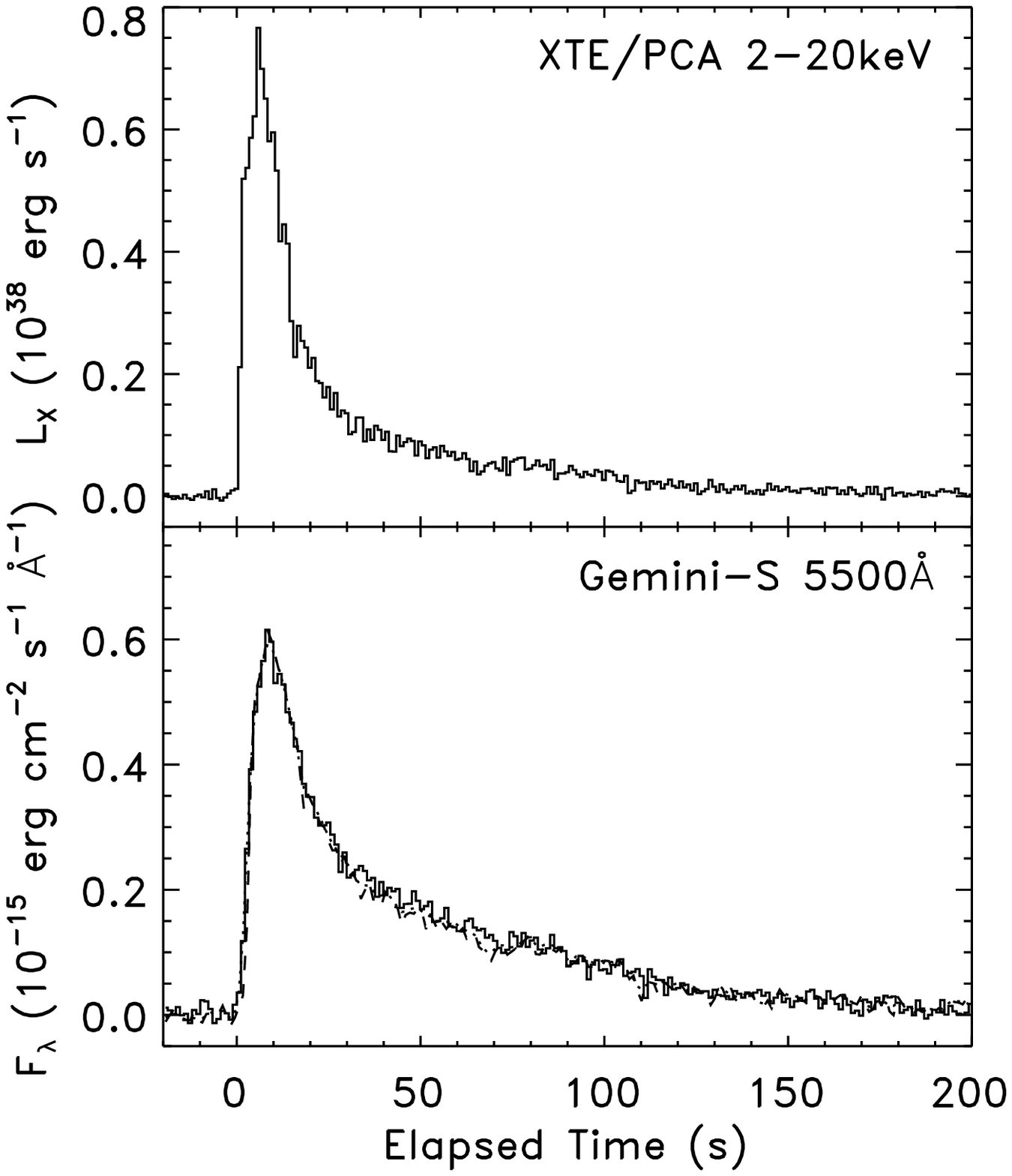}
\caption{Single black body fits to MB1.  The underlying non-burst
X-ray luminosity that has been subtracted off is
$1.6\times10^{36}$\,erg\,s$^{-1}$.  The dotted line corresponds to the
formal best fit, MB1-1 in Table~\ref{GaussFitTable}.  The dashed line
corresponds to the broad secondary maximum which we suspect may be the
true solution, MB1-2.  Note that the latter fits the rise time of the
optical burst better.}
\label{MBOneFig}
\end{figure}

We have left MB1 until last as this is the least constrained case.  We
do not have UV data allowing independent fits of $T_{\rm min}$ and
$T_{\rm max}$, but neither are we well justified in assuming
similarity to MB2 or MB3.  MB1 occurred on the preceding night, and
also at phase 0.2 when our visibility of the system was significantly
different.  We might then expect any of the parameters to be
different.  Nonetheless, we will begin from the same
starting point as for MB3, i.e.\ assuming that $T_{\rm min}=18500$\,K,
as for MB2, and performing an unconstrained fit on the other
parameters.  This burst is more problematic than the preceding
ones, as there are actually two plausible solutions differing mainly
in the width of the transfer function.  We show in
Fig.~\ref{WidthChiFig} the dependence of normalized $\chi^2$ on the
Gaussian width, $\sigma_{\tau}$ for all three bursts.  For MB2 and MB3
there was a single, well-defined minimum, and the location is
approximately the same for these two bursts.  This width is clearly
not consistent with the data from MB1, and a narrower transfer
function is required.  Formally the best fit occurs for a very narrow
transfer function: MB1-1 as listed in Table~\ref{GaussFitTable}.
$\chi^2$ is a rather complex function of $\sigma_{\tau}$ is this
region, and the best fit represents a very narrow minimum (hence the
unrealistically small uncertainties for $\sigma_{\tau}$ in model MB1-1).
These characteristics suggest that the behavior at small
$\sigma_{\tau}$ may reflect random correlations in the noise between
X-ray and optical lightcurves.  A secondary minimum is present
corresponding to a broader transfer function, model MB1-2.  This minimum
more closely resembles those seen for MB2 and MB3, and we suspect that
this could be the true solution.  In support of this, we show both
models fitted to the data in Fig.~\ref{MBOneFig}.  Model MB1-1 predicts a
steeper optical rise than observed, whereas MB1-2 gives
approximately the correct rise time.  It is not obvious in what
respect MB1-1 is preferred, also indicating that it could owe more
to fitting the noise than to fitting the real structure.  We will
therefore adopt MB1-2 as the preferred fit to this burst.

The parameters obtained do seem plausible.  We derive a
shorter lag and narrower lag distribution (independently of whether we
adopt MB1-1 or MB1-2).  Both are to be expected at
phase 0.2, as the companion star will make less contribution, and will
do so at a shorter lag.  The temperature and area derived are
comparable, although a smaller reprocessing area is found, again as
expected at this phase.  Based on these parameters there is no need to
invoke significant differences in the geometry or temperature of the
reprocessing regions between the two nights.

Nonetheless we did attempt fits with a range of $T_{\rm min}$ values.
As for other bursts we find that formally the best fits are for
unphysically low base temperatures ($T_{\rm min} \la 4000$\,K, but
that a large range of values larger than this also give acceptable
fits.  There is no reason to prefer a different $T_{\rm min}$ value to
MB2 or MB3.

%
%
\section{The Ultraviolet Burst Spectrum}

\begin{figure}
\epsfig{width=3.5in,file=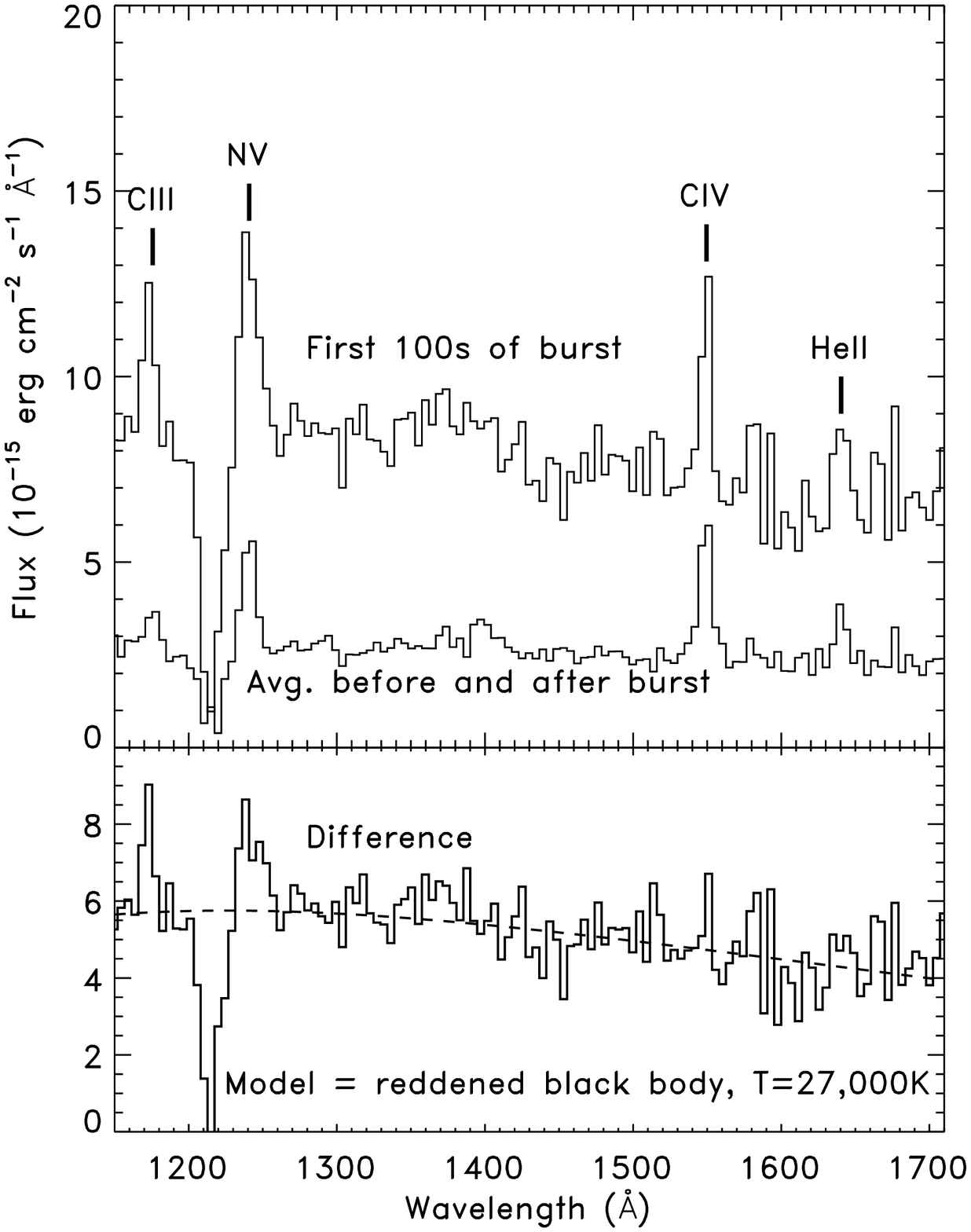}
\caption{Far-UV spectrum of MB2.  The upper panel shows the spectrum
  of the first 100\,s of the burst compared to that before and after.
  The lower panel shows the difference between the two.  The dashed
  line shown in the lower panel is the expected black body burst
  spectrum based on the time-averaged temperature evolution fitted
  earlier.  All spectra shown here have been rebinned to 4.7\,\AA\ per
  pixel.}
\label{BurstSpecFig}
\end{figure}

As our MB2 UV data were obtained in a spectroscopic mode, as well as
obtaining the lightcurve of the reprocessed burst, we have a unique
opportunity to examine its spectrum.  Fig.~\ref{BurstSpecFig} shows
the spectrum of the extra light during the first 100\,s of the burst.
This was constructed by extracting a series of 100\,s sub-spectra
using the {\sc stsdas} task {\sc inttag} and processing them in the
same way as regular spectra.  The non-burst spectrum was defined from
300\,s intervals before and after the burst.  More details of the
spectral reduction will be provided in Paper II.

The burst spectrum appears to be continuum dominated with only a weak
contribution from the lines.  C\,{\sc iii} 1175\,\AA\ is particularly
strongly enhanced, and to a lesser extent N\,{\sc v} 1240\,\AA, but
neither dominate the flux.  The burst is not pronounced at all in
C\,{\sc iv} 1550\,\AA\ or He\,{\sc ii} 1640\,\AA.

The continuum shape is consistent with the burst models described
earlier, being well fitted by the difference between a black body with
temperature $\sim27,000$\,K and one at $\sim18,500$\,K.  The upper
temperature corresponds to the expected flux-weighted average
temperature during the 100\,s interval contributing to the burst as
calculated from model 1 for MB2.  This shows that the model does
reproduce the wavelength dependence as well as the temporal
evolution.  It also indirectly supports our low reddening value
($\ebv=0.06\pm0.03$) derived from the 2175\,\AA\ interstellar feature
(Paper III).  If this reddening had been underestimated due to an
anomalously weak 2175\,\AA\ feature then in general we would not
expect the temperature implied by the shape of the spectrum to be
consistent with that earlier obtained only from time-dependence of the
reprocessed light, and the far-UV spectrum would be redder.

%
\section{Discussion}

\subsection{Evidence for phase dependence of the burst response}
\label{EchoTomSection}
One of the primary goals of echo mapping is echo-tomography:
phase-dependent echo maps in which different components, such as the
disk and companion star, can be disentangled based on different
amplitudes and phasings of phase dependence.  In principle this
technique could even be used to obtain a direct measure of the binary
separation and orbital inclination independent of other constraints.
Real datasets have, however, fallen far short of these aspirations,
and we certainly would not claim that the results presented here
fulfill the hopes of echo-mapping.  Nonetheless, we do see changes in
the response, and it is worthwhile to investigate if they are
consistent with expected phase-dependent changes, and can constrain
the parameters of \uyvol.

One of the major limitations, of course, is that MB1 was obtained on a
different night to MB2 and MB3.  Since MB1 provides the main
sensitivity to phase-dependent changes, we must apply the caveat that
the differences observed could reflect secular changes in the actual
reprocessing geometry rather than just differences in our viewing
angle.  The persistent (pre-burst) X-ray flux is essentially the same
for MB1 and MB2.  It is lower for MB3, but there is some evidence
for dipping at this time.  This implies no substantial difference in
the accretion rate between the MB1 and MB2 epochs.  Optical
lightcurves will be compared in Paper III and should be very sensitive
to changes in the reprocessing geometry between epochs.  The optical
lightcurve at the time of MB1 is somewhat different, but the
differences are not dramatic and may primarily reflect different
realizations of the strong flickering which is present in the optical
data.  The amplitude of the differences between MB1 and MB2 epochs
($\la25$\,\%) is comparable to that between MB2 and MB3 epochs, and
also to the amplitude of individual flickering events.  There is thus
no compelling evidence for a change in the state of the system, but
this possibility cannot be securely discounted.

\begin{figure}
\epsfig{angle=90,width=3.5in,file=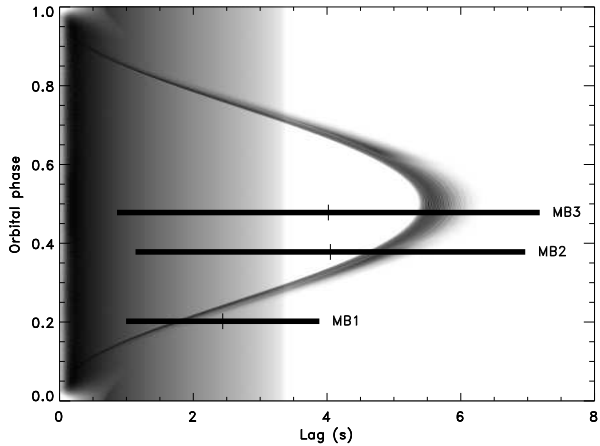}
\caption{Predicted echo tomogram calculated according to
  \citet{OBrien:2002a}.  The broad non-varying region to the left is
  the disk, the moving component is the irradiated face of the
  companion star.  Solid bars indicate the response inferred from
  Gaussian fitting of MB1 (lower), MB2 (middle), and MB3a and b
  combined (upper).  The bar represents the lags over
  which the Gaussian model is above half response (i.e.\ its length is
  the full-width at half maximum).}
\label{EchotomFig}
\end{figure}

Assuming it is reasonable to compare the responses from the three
bursts, the major difference is that MB2 and MB3 both have responses
that are lagged and smeared by more.  This is as expected from their
phases.  Fig.~\ref{EchotomFig} shows predicted responses as a function
of orbital phase calculated using the methods described by
\citet{OBrien:2002a} for model 2 ($q=0.2$).  Given the limitations
inherent in approximating the true response with a Gaussian, the range
of lags observed is in reasonable agreement with predictions.  In
particular, we do expect that MB2 (phase 0.38) will have a similar lag
distribution to MB3 (phase $\sim 0.5$), whereas we expect a quicker
and less smeared response from MB1, as observed.  The amplitude of the
difference between responses is in approximate agreement with
expectations.  When plotted in this way, it is apparent that all
bursts appear to show the onset of a strong response at approximately
the same time, but that MB2 and MB3 extend for longer beyond that.

While the general picture is in agreement with predictions, it is
clear that the data are not precise enough to constrain the system
parameters further.  Nonetheless, the very fact that the response
appears to change with phase, and with such a large amplitude, does
indicate significant response from the companion, and the large
smearing of the response around phase 0.5 indicates that the disk must
also contribute.

We can also consider these results in the context of other
observations of this and other sources.  The comparison of most
interest is with \citet{Schoembs:1990a}.  These authors found that
optical bursts (with no simultaneous X-ray coverage) exhibited longer
rise times when the companion star would be on the far side of the
disk.  This is equivalent to saying that the smearing increases close
to phase 0.5 (as we also find), provided that the intrinsic rise time
of the X-ray bursts is the same.  The large dispersion in the response
that we see close to phase 0.5 is also not unprecedented.  Similar
dispersions were found by \citet{Truemper:1985a} for 4U\,1636--536,
and by \citet{Kong:2000a} for GS\,1826--24.  In all of these cases
they likely indicate contribution from both disk and companion, as a
response from the companion alone will vary in lag, but should always
have a relatively narrow response.

\subsection{Local diffusion time delays}
\label{DiffusionSection}

There are two likely sources of smearing of the reprocessed signal.
The delays most commonly discussed (and considered so far) are those
from light-travel times across the binary; global delays.  A finite
reprocessing time could also introduce a local delay.  X-rays above
about 1\,keV will experience a low-enough photo-electric absorption
opacity to deposit energy at a significant optical depth in an
atmosphere.  Reprocessed energy will then have to diffuse outward,
and so there will be an additional diffusion time delay.

This topic was considered initially by \citet{Pedersen:1982a} in
analyzing bursts from 4U~1636--536.  They estimated that the diffusion
time for a typical photon would be $\sim0.6$\,s, largely negligible
compared to the expected light travel time delays.
\citet{Cominsky:1987a} examined the problem more rigorously,
calculating time-dependent responses of a hot stellar atmosphere to an
X-ray burst, including the effect of the burst on the atmospheric
temperature structure and opacities.  Their results were in agreement
with those of \citet{Pedersen:1982a}, and they found that 50\,\%\ of
the reprocessed light is expected within just 0.2\,s, but that there
was also a very extended tail to the response up to 10\,s.  Finally,
\citet{McGowan:2003a} also considered this issue, but neglected
photo-absorption, which will be critical for the soft X-rays produced
in a burst.  Based on these calculations, we would not expect our
method to be significantly affected by reprocessing times, as a
Gaussian response approximation will be dominated by when most of the
light emerges, and this occurs within a fraction of a second.
Indeed, our observations are broadly consistent with expectations from
light travel time delays and, allowing for the limitations of the
Gaussian approximation, do not obviously require additional local
delays.

This conclusion may be different from that obtained for correlated
flickering in the low-mass X-ray binaries Sco X-1 and LMC X-2, for
which \citet{McGowan:2003a} argue that lags may be too long for light
travel times alone.  More and better data are needed to confirm this
discrepancy, but if confirmed the key difference may be in the nature
and spectrum of the X-ray irradiation.  \citet{Cominsky:1987a} also
considered harder irradiation than that provided by an X-ray burst,
and found that it could increase the diffusion timescale significantly
(see also \citet{McGowan:2003a}).  Thus while reprocessing times are
of marginal significance in considering burst reprocessing, they may
be of more importance for other applications where further
investigation is needed.  The corollary of this is that the relatively
soft irradiation of bursts may provide the least biased way to isolate
light travel time delays and hence the ideal signal for echo-tomography.

\subsection{Expected temperatures}

It is interesting to compare the deduced reprocessing temperatures
with those expected based on the X-ray luminosity, as this provides a
test of the reprocessing efficiency.  We will consider MB2, for which
the temperatures are best constrained.  For this burst, the persistent
luminosity (at 6.8\,kpc) was $2.1\times10^{36}$\,erg\,s$^{-1}$ and the
peak burst luminosity was $6.9\times10^{37}$\,erg\,s$^{-1}$.  

Irradiation of the companion star is sensitive to the mass ratio, as
smaller companions will more readily be shielded by the disk.  For
example, if we take model 1 ($q=0.08$) and the typical effective disk
opening angle of 12$^{\circ}$ deduced by \citet{deJong:1996a} then the
companion would be completely shielded.  This is clearly not the case,
as the companion can directly eclipse the neutron star, but we cannot
quantify how directly it is illuminated.  Note that the opening angle
need not be that of the disk rim, but could represent material in the
inner disk that can shield the companion.  For models 2 and 3,
calculations are more straightforward and the differences between the
two are less dramatic.  For a disk opening angle of 6$^{\circ}$ we
expect angles of incidence of greater than 45$^{\circ}$ from normal,
increasing to 60$^{\circ}$ for opening angles of 12$^{\circ}$.  We
assume an X-ray albedo for the companion of $\sim0.4$ following
\citet{deJong:1996a}.  We then expect the highest persistent irradiation
temperature on the companion of 16,000--21,000\,K depending on mass
ratio (for the range 0.20--0.34) and disk opening angle (in the range
6--12$^{\circ}$).  This is the temperature at the point closest to the
compact object and most of the irradiated region is at
temperatures less than this as the angle of incidence is steeper.  At
the burst peak we would expect temperatures of 38,000--50,000\,K.

For the disk we estimate the estimated temperature at the disk rim
using the semi-empirical prescription of \citet{Dubus:1999a} and
\citet{Dubus:2001a}.  Assuming Dubus' estimate of the irradiation
efficiency and a disk radius of $5\times10^{10}$\,cm we expect a
persistent irradiation temperature of 9000\,K and a peak burst
temperature of 21,000\,K.  Both of these are the lowest temperatures
in the disk, and most of the disk area is hotter than this (the
opposite case to that calculated for the companion star).  For
comparison, the persistent luminosity of
$2.1\times10^{36}$\,erg\,s$^{-1}$ corresponds to a mass transfer rate
of $\sim1.2\times10^{16}$\,g\,s$^{-1}$.  For a steady state
lobe-filling disk the effective temperature due to viscous heating
should be just 2500\,K, indicating that the disk is in a regime that
can only remain in a high state with irradiative heating.

Our earlier conclusions that the reprocessed bursts likely arise from
a combination of emission from the companion star and disk are
consistent with these calculations.  Our estimate of the peak
reprocessing temperature is 35,000\,K.  The irradiated area of the
companion star is expected to have temperatures possibly extending 
up to 50,000\,K, while the disk could plausibly have temperatures from
21,000\,K upward.  The temperature ranges are overlapping, with the
greater distance to the companion offset by its more direct
illumination, and are consistent with our observations.  

%
%
\section{Conclusions}

We have reported several simultaneous X-ray bursts in the low-mass
X-ray binary \uyvol, including the highest quality reprocessed optical
coverage, and the only reprocessed UV burst that we are aware of in
any source.  These results allow a more thorough test of the paradigm
for reprocessing of X-ray bursts, and reprocessing of X-ray emission
in general, than previously possible.  Several key assertions about
reprocessing have been tested.

i) X-rays are absorbed at relatively high optical depths, thermalized,
and re-emitted with a quasi-black body spectrum.  By obtaining the UV
spectrum of the extra light produced during a burst, we have shown
that it is indeed dominated by continuum emission, and that the shape
of that continuum is consistent with black body emission with
temperatures as inferred from the lightcurves of the bursts.  The
discrepancy with the optical flux, however, suggests that a
single-temperature black body model is not sufficient.

ii) Reprocessed emission in the optical and UV is lagged predominantly
by light travel times rather than by local photon diffusion.  This
appears to be borne out by our observations as light travel times are
sufficient to explain the lags and smearing observed.  Note that this
conclusion (and the preceding one) may not be true in general, but may
be specific to irradiation with the spectrum of an X-ray burst.
Harder or softer irradiation is likely to produce a different response.

iii) X-rays are reprocessed both by the accretion disk and by the
companion star.  This is supported by our observations, as we do
appear to see a phase-dependence of both the lag and the smearing of
the response.  If the response came from the disk alone, we would
expect little variation.  If it came from the companion alone, we
predict variations in the lag, but not in the smearing.  The
temperatures that both the disk and companion are expected to reach
during a burst are consistent with our observations.  The failure of a
single-zone reprocessor model to explain the relative UV and optical
fluxes in the burst also supports a multi-component response.  Our
observations do not appear to resolve distinct lags from disk and
companion, however, possibly a consequence of the long bursts
exhibited by \uyvol.  Sources which exhibit short duration bursts may
be better suited to this analysis.

%
\acknowledgments

This work includes observations with the NASA/ESA {\it Hubble Space
Telescope}, obtained at STScI, which is operated by AURA Inc.\ under
NASA contract No.\ NAS5-26555.  Support for \HST\ proposal GO\,9398
was provided by NASA through a grant from STScI.  RIH also
acknowledges support by NASA through Hubble Fellowship grant
\#HF-01150.01-A awarded by STScI.  This work uses observations
obtained at the Gemini Observatory, which is operated by the AURA
Inc., under a cooperative agreement with the NSF on behalf of the
Gemini partnership: the National Science Foundation (United States),
the Particle Physics and Astronomy Research Council (United Kingdom),
the National Research Council (Canada), CONICYT (Chile), the
Australian Research Council (Australia), CNPq (Brazil) and CONICET
(Argentina).  We would like to thank Rodrigo Carrasco and the other
staff involved in obtaining these data for putting in the extra effort
to make this unusual observation a success!  RIH would like to thank
Danny Steeghs for obtaining prior imaging of the field to aid our
\HST\ preparation.  This work has made use of the NASA Astrophysics
Data System Abstract Service.

\clearpage





\begin{thebibliography}{}

\bibitem[Chabrier \& Baraffe(2000)]{Chabrier:2000a} Chabrier, G., \& 
Baraffe, I.\ 2000, \araa, 38, 337 

\bibitem[Cominsky, London, \& Klein(1987)]{Cominsky:1987a} Cominsky, 
L.~R., London, R.~A., \& Klein, R.~I.\ 1987, \apj, 315, 162 

\bibitem[de Jong, van Paradijs, \& Augusteijn(1996)]{deJong:1996a} 
de Jong, J.~A., van Paradijs, J., \& Augusteijn, T.\ 1996, \aap, 314,
484 

\bibitem[Dubus et al.(1999)]{Dubus:1999a} Dubus, G., Lasota, J.-P., 
Hameury, J.-M., \& Charles, P.\ 1999, \mnras, 303, 139 

\bibitem[Dubus et al.(2001)]{Dubus:2001a} Dubus, G., Hameury, 
J.-M., \& Lasota, J.-P.\ 2001, \aap, 373, 251 
 
\bibitem[Fitzpatrick(1999)]{Fitzpatrick:1999a} Fitzpatrick, E.~L.\ 1999, 
\pasp, 111, 63 

\bibitem[Fukugita, Shimasaku, \& Ichikawa(1995)]{Fukugita:1995a} 
Fukugita, M., Shimasaku, K., \& Ichikawa, T.\ 1995, \pasp, 107, 945 

\bibitem[Gaskell \& Peterson(1987)]{Gaskell:1987a}
         Gaskell, C.~M., Peterson, B.~M. 1987, \apjs, 65, 1

\bibitem[Gottwald et al.(1986)]{Gottwald:1986a} Gottwald, M., Haberl, 
F., Parmar, A.~N., \& White, N.~E.\ 1986, \apj, 308, 213 

\bibitem[Haswell et al.(2002)]{Haswell:2002a} Haswell, C.~A., Hynes, 
R.~I., King, A.~R., \& Schenker, K.\ 2002, \mnras, 332, 928 

\bibitem[Horne(1994)]{Horne:1994a} Horne, K.\ 1994, Astronomical 
Society of the Pacific Conference Series, 69, 23 

\bibitem[Horne et al.(1991)]{Horne:1991a} Horne, K., Welsh, W.~F., 
\& Peterson, B.~M.\ 1991, \apjl, 367, L5 

\bibitem[Hynes et al.(1998)]{Hynes:1998b} Hynes, R.~I., O'Brien, 
K., Horne, K., Chen, W., \& Haswell, C.~A.\ 1998, \mnras, 299, L37 

\bibitem[Hynes et al.(2002)]{Hynes:2002a} Hynes, R.~I., Haswell, 
C.~A., Chaty, S., Shrader, C.~R., \& Cui, W.\ 2002, \mnras, 331, 169 

\bibitem[Hynes et al.(2003)]{Hynes:2003a}
        Hynes, R.~I., Haswell, C.~A., Cui, W., Shrader, C. ~., O'Brien, K.,
        Chaty, S., Skillman, D.~R., Patterson, J., Horne, K.\ 2003, \mnras,
        345, 292 

\bibitem[Hynes(2005)]{Hynes:2004a}
Hynes, R.~I., in ``The Astrophysics of Cataclysmic Variables
and Related Objects'', ASP Conf.\ Ser.\ Vol.\ 330, Eds.\ J.~M.~Hameury \&
J.~P.~Lasota, p237

\bibitem[Jahoda et al.(2006)]{Jahoda:2005a}
Jahoda, K., Markwardt, C.~B., Radeva, Y., Rots, A.~H., Stark, M.~J.,
Swank, J.~H., Strohmayer, T.~E., Zhang, W., 2006, \apjs, in press

\bibitem[Jonker \& Nelemans(2004)]{Jonker:2004a} Jonker, P.~G., \& 
Nelemans, G.\ 2004, \mnras, 354, 355 

\bibitem[Jonker et al.(2005)]{Jonker:2005a} Jonker, P.~G., Steeghs, 
D., Nelemans, G., \& van der Klis, M.\ 2005, \mnras, 356, 621 

\bibitem[Koen(2003)]{Koen:2003a} Koen, C.\ 2003, \mnras, 344,
  798 

\bibitem[Kong et al.(2000)]{Kong:2000a} Kong, A.~K.~H., Homer, L., 
Kuulkers, E., Charles, P.~A., \& Smale, A.~P.\ 2000, \mnras, 311, 405 

\bibitem[Kuulkers et al.(2003)]{Kuulkers:2003a} Kuulkers, E., den 
Hartog, P.~R., in't Zand, J.~J.~M., Verbunt, F.~W.~M., Harris, W.~E., \& 
Cocchi, M.\ 2003, \aap, 399, 663 

\bibitem[Landolt(1992)]{Landolt:1992a} Landolt, A.~U.\ 1992, \aj, 
104, 340 

\bibitem[Lawrence et al.(1983)]{Lawrence:1983a} Lawrence, A., et al.\ 
1983, \apj, 271, 793 

\bibitem[Liu et al.(2001)]{Liu:2001a} Liu, Q.~Z., van Paradijs, 
J., \& van den Heuvel, E.~P.~J.\ 2001, \aap, 368, 1021 

\bibitem[McGowan et al.(2003)]{McGowan:2003a} 
McGowan, K.~E., Charles, P.~A., O'Donoghue, D., \& Smale, A.~P.\ 2003, 
\mnras, 345, 1039 

\bibitem[Naylor(1998)]{Naylor:1998a} Naylor, T.\ 1998, \mnras, 296, 
339 

\bibitem[O'Brien et al.(2002)]{OBrien:2002a} O'Brien, K., Horne, K., 
Hynes, R.~I., Chen, W., Haswell, C.~A., \& Still, M.~D.\ 2002, \mnras, 334, 
426 

\bibitem[Parmar et al.(1985)]{Parmar:1985a} Parmar, A.~N., White, 
N.~E., Giommi, P., Haberl, F., Pedersen, H., \& Mayor, M.\ 1985, \iaucirc, 
4039

\bibitem[Parmar et al.(1986)]{Parmar:1986a} Parmar, A.~N., White, 
N.~E., Giommi, P., \& Gottwald, M.\ 1986, \apj, 308, 199 

\bibitem[Pearson et al.(2006)]{Pearson:2006a} Pearson, K.~J., et al.\
  2006, \apj, submitted (Paper II)

\bibitem[Pedersen et al.(1982)]{Pedersen:1982a} Pedersen, H., et al.\ 
1982, \apj, 263, 325 

\bibitem[Press et al.(1992)]{Press:1992a}
Press, W.~H., Teukolsky, S.~A., Vetterling, W.~T. \& Flannery, B.~P.\ 1992,
Numerical Recipes in C: The Art of Scientific Computing, 2nd Edn.,
CUP, 1992, p408

\bibitem[Profitt et al.(2002)]{Profitt:2002a}
  Profitt, C., et al.\ 2002, STIS Instrument Handbook, Version
  6.0, STScI, Baltimore

\bibitem[Schenker \& King(2002)]{Schenker:2002a} Schenker, K., \& 
King, A.~R.\ 2002, ASP Conf.~Ser.~261: The Physics of Cataclysmic Variables 
and Related Objects, 261, 242 

\bibitem[Schoembs \& Zoeschinger(1990)]{Schoembs:1990a} Schoembs, 
R.~\& Zoeschinger, G.\ 1990, \aap, 227, 105

\bibitem[Shahbaz et al.(2004)]{Shahbaz:2004a} Shahbaz, T., Casares, 
J., Watson, C.~A., Charles, P.~A., Hynes, R.~I., Shih, S.~C., \& Steeghs, 
D.\ 2004, \apjl, 616, L123 

\bibitem[Thorsett \& Chakrabarty(1999)]{Thorsett:1999a} Thorsett, 
S.~E., \& Chakrabarty, D.\ 1999, \apj, 512, 288 

\bibitem[Truemper et al.(1985)]{Truemper:1985a} Truemper, J., Sztajno, 
M., Pietsch, W., van Paradijs, J., \& Lewin, W.~H.~G.\ 1985, Space Science 
Reviews, 40, 255 
%
\bibitem[van Paradijs, van der Klis, \&
Pedersen(1988)]{vanParadijs:1988a} van Paradijs, J., van der Klis, M.,
\& Pedersen, H.\ 1988, \aaps, 76, 185

\bibitem[Wade et al.(1985)]{Wade:1985a} Wade, R.~A., Quintana, H., 
Horne, K., \& Marsh, T.~R.\ 1985, \pasp, 97, 1092 

\bibitem[White \& Peterson(1994)]{White:1994a}
White, R.~J., Peterson, B.~M. 1994, \pasp, 106, 879

\bibitem[Wolff et al.(2002)]{Wolff:2002a} Wolff, M.~T., Hertz, P., 
Wood, K.~S., Ray, P.~S., \& Bandyopadhyay, R.~M.\ 2002, \apj, 575, 384 

\bibitem[Wolff et al.(2005)]{Wolff:2005a} Wolff, M.~T., Becker, 
P.~A., Ray, P.~S., \& Wood, K.~S.\ 2005, \apj, 632, 1099 
 
\end{thebibliography}
\end{document}